\documentclass[twocolumn]{aastex631}

\usepackage{graphicx}
\usepackage[caption=false]{subfig}
\usepackage[T1]{fontenc}
\usepackage{natbib}
\usepackage{amsmath}
\usepackage{float}

\begin{document}

\title{Simulating ULXs and blazars as GRMHD accretion flows around a black hole}

\author[0000-0002-4834-4211]{Mayank Pathak}
\affiliation{Joint Astronomy Programme, Department of Physics, Indian Institute of Science, Bengaluru 560012, India}

\author[0000-0002-3020-9513]{Banibrata Mukhopadhyay}
\affiliation{Joint Astronomy Programme, Department of Physics, Indian Institute of Science, Bengaluru 560012, India}
\affiliation{Department of Physics, Indian Institute of Science, Bengaluru 560012, India}

\begin{abstract}

General relativistic magnetohydrodynamic (GRMHD) simulations have been instrumental in our understanding of high energy astrophysical phenomena over the past two decades. Their robustness and modularity make them a great tool for understanding the dynamics of various astrophysical objects. In this paper we have used GRMHD simulations to understand the accretion flows of ultraluminous X-ray sources (ULXs) and blazars. ULXs are enigmatic sources which exhibit very high luminosities (super-Eddington for stellar mass black holes) even in their low-hard state. Numerical steady state calculations have shown that this behaviour can be explained by considering ULXs to be highly magnetised advective accretion sources around stellar-mass black holes. Our simulation confirms that such an accretion flow can indeed produce the high luminosities observed in ULXs. Further to continue towards the supermassive black holes, we have also modeled blazars and have used our simulation results to explain the apparent dichotomy in the two blazar classes: flat spectrum radio quasars (FSRQs) and BL Lacertae (BL Lacs). Our results show that FSRQ and BL Lacs show different spectral characteristics due to a difference in their magnetic field characteristics. The different categories of FSRQs and BL Lacs have also been explained by the interplay between the spin, magnetic field and accretion rate of the central supermassive black hole.

\end{abstract}

\keywords{Blazars (164) --- High energy astrophysics (739) --- Gravitation (661) --- Relativistic jets (1390) --- Stellar accretion disks (1579) --- Magnetohydrodynamical simulations (1966) --- Ultraluminous x-ray sources (2164)}

\section{Introduction} \label{sec:intro}

Accretion flows around black holes serve as natural laboratories for studying various high energy physical phenomena. Analysing the observations from these systems has led to insights into physics under extreme conditions that are still largely inaccessible in terrestrial experiments. To study the properties of the underlying accretion flow, in addition to astronomical observations, detailed numerical general relativistic magnetohydrodynamic (GRMHD) simulations \citep{harm} have been used \citep{eht1}.

In the last twenty years, GRMHD simulations have developed into highly modular and sophisticated tools for investigating the behaviour of magnetised plasma in the presence of strong gravitational fields, particularly near compact objects such as black holes and neutron stars. These simulations have been instrumental in enhancing our understanding of accretion physics and corresponding outflow mechanisms such as jet formation, winds, etc. \citep{rn2022,rn2011}.

On the other hand, in a series of papers from 2019-2020, Mondal and Mukhopadhyay explored the role of strong magnetic fields in the overall dynamics of accretion flows around black holes by using numerical steady state solutions. Using their results, they were able to explain the observational properties of hard state ultraluminous X-ray sources (ULXs) (\citealt{ULX19,ULX20}; ULX19 hereafter) and the apparent dichotomy in the observed luminosities of FSRQ and BL Lac objects (\citealt{BL19}; BL19 hereafter). 

In this paper, we have considered GRMHD simulations to verify Mondal and Mukhopadhyay's results and further validate their findings. Our particular emphasis is on two cases, described below.

\subsection{Ultraluminous X-ray sources in hard state}
Many X-ray sources exhibit cyclic transitions between high-soft and low-hard spectral states (e.g. \citealt{bell02}). As the names of these states suggest, the high-soft state is characterised by high luminosities and a soft/thermal spectrum, while the low-hard state has low luminosities and a hard/non-thermal spectrum \citep{bell10}.

However, certain ULXs peculiarly show high luminosities even in their hard, power law dominated spectral state (see, e.g., \citealt{sutton}). These sources are non-nuclear, point-like and very rare. Considering isotropic emissions, their luminosities lie within the range of $3\times 10^{39} - 3\times 10^{41}$ ergs/s, which are super-Eddington luminosities for stellar mass sources, where Eddington luminosity is defined as 
\begin{equation}
L_{edd}=\frac{4\pi cGM}{\kappa_{es}},
\end{equation}
with $G$ being the gravitation constant, $c$ the speed of light, $M$ the mass of the source and $\kappa_{es}$ the electron scattering opacity \citep{fab06}.

On the other hand, Cyg X-1 is a frequently studied (non ULX) persistent X-ray binary due to its intense brightness and intricate behaviour.
It harbours a $21.2M_\odot$ black hole \citep{cygm}. It also remains in a power law dominated hard spectral state for most of its time ($\sim 90\%$) with an exponential cutoff at 150 keV \citep{cyg1,cyg3}. This hard state is originated due to non-thermal emissions, indicating an optically thin, possibly low density flow. Advective, sub-Keplerian, sub-Eddington accretion flows are known to be optically thin and have been used to understand such spectral characteristics in accreting sources.
Also, polarization studies of Cyg X-1 have revealed high magnetic field strengths ($\sim 10^6$ G) at the base of its jet \citep{cyg2}. 
These show the evidences for strong magnetic fields in black hole X-ray binaries in hard states. Such flows need to be modeled by highly magnetised sub-Eddington advective accretion flows.

In ULX19, Mondal and Mukhopadhyay showed by numerical calculations that the peculiar super-Eddington luminosities in hard state ULXs can be explained by considering them to be highly magnetised sub-Eddington advective accretion flows around stellar mass black holes. This eliminates the need for intermediate mass black holes for explaining these high luminosities. The sub-Eddington accretion rate also explains the hard state spectral signatures.

\subsection{FSRQ/BL Lac dichotomy}

One of the most distinctive features of radio-loud active galactic nuclei (AGNs) is the presence of large scale relativistic jets. The unified models of AGNs explain the various kinds of observed AGNs as the same object observed along different lines of sight (LOS) \citep{uni1,uni2}. Depending on its angle, as the LOS passes through dust and clouds around the AGN, different kinds of characteristics are observed which can be used to divide AGNs into different classes. One such class of AGNs is called blazars, in which the jets are along the LOS of the observer. 

Based on their spectral characteristics, blazars can be broadly classified into two categories, namely, flat spectrum radio quasars (FSRQs) and BL Lacertae (BL Lacs). The equivalent width of the optical emission line is $\geq5\text{\r{A}}$ for FSRQs while it is $<5\text{\r{A}}$ for BL Lacs \citep{uni2}.

Blazar spectra have two distinct peaks, lying in the optical-UV to $\gamma$-ray regime. The low energy peak is well explained by synchrotron emission produced by electrons moving at relativistic speeds. The $\gamma$-ray peak is formed due to the inverse Compton (IC) scattering of photons by higher energy electrons in the jet plasma. In the case of BL Lacs, these photons are provided by synchrotron emission. This process is called synchrotron self-Comptonization (SSC). For FSRQs however, the IC scattering is produced by SSC and external Comptonization (EC) both \citep{der93,bla00,sik09}. In addition to this, BL Lac objects are further classified into three classes based on the location of their synchrotron peak on the frequency axis ($\nu_{sy}$) in their spectra in the observer rest frame. These classes are: low-synchrotron peak BL Lacs (LSP BL Lacs, $\nu_{sy}<10^{14}$ Hz), intermediate-synchrotron peak BL Lacs (ISP BL Lacs, $10^{14}$ Hz $<\nu_{sy}<10^{15}$ Hz) and high-synchrotron peak BL Lacs (HSP BL Lacs, $\nu_{sy}>10^{15}$ Hz) \citep{fermi10}.   

Several attempts have been made to understand the apparent dichotomy in the observed FSRQ/BL Lac luminosities. These differences have been argued to be a product of the intrinsic properties of the system along the lines of the unification model of AGNs. In BL19, Mondal and Mukhopadhyay showed by analysing the observed blazar data from the $Fermi$ catalogue \citep{fermi10} that, while FSRQs have higher observed luminosities than BL Lac sources, upon debeaming, FSRQs show a strong dependence on the amount of EC suffered by the incoming photons.

In this paper, we have used GRMHD simulations to explore highly magnetised advective accretion flows around rotating black holes, to understand various magnetised sources. We study the variation of outflow power with respect to various system parameters. Our simulation results show that the power derived from highly magnetised sub-Eddington advective flows around rapidly rotating stellar mass black holes is within the observed ULX luminosity range. We have also explained the apparent dichotomy in FSRQ/BL Lac observations as a result of different magnetic field characteristics of their respective underlying accretion flows.

The paper is organised as follows. In section 2, we desribe the simulation setup used for our study. In section 3, we discuss the results of our simulations and show the obtained outflow power and magnetic field profiles. In section 4, we discuss the outflow power profiles and magnetic field stress profiles in the context of FSRQ/ BL Lac dichotomy. We conclude in section 5 with the discussion of future work needed to be done. 

\begin{figure*}
\centering
\includegraphics[width=\textwidth]{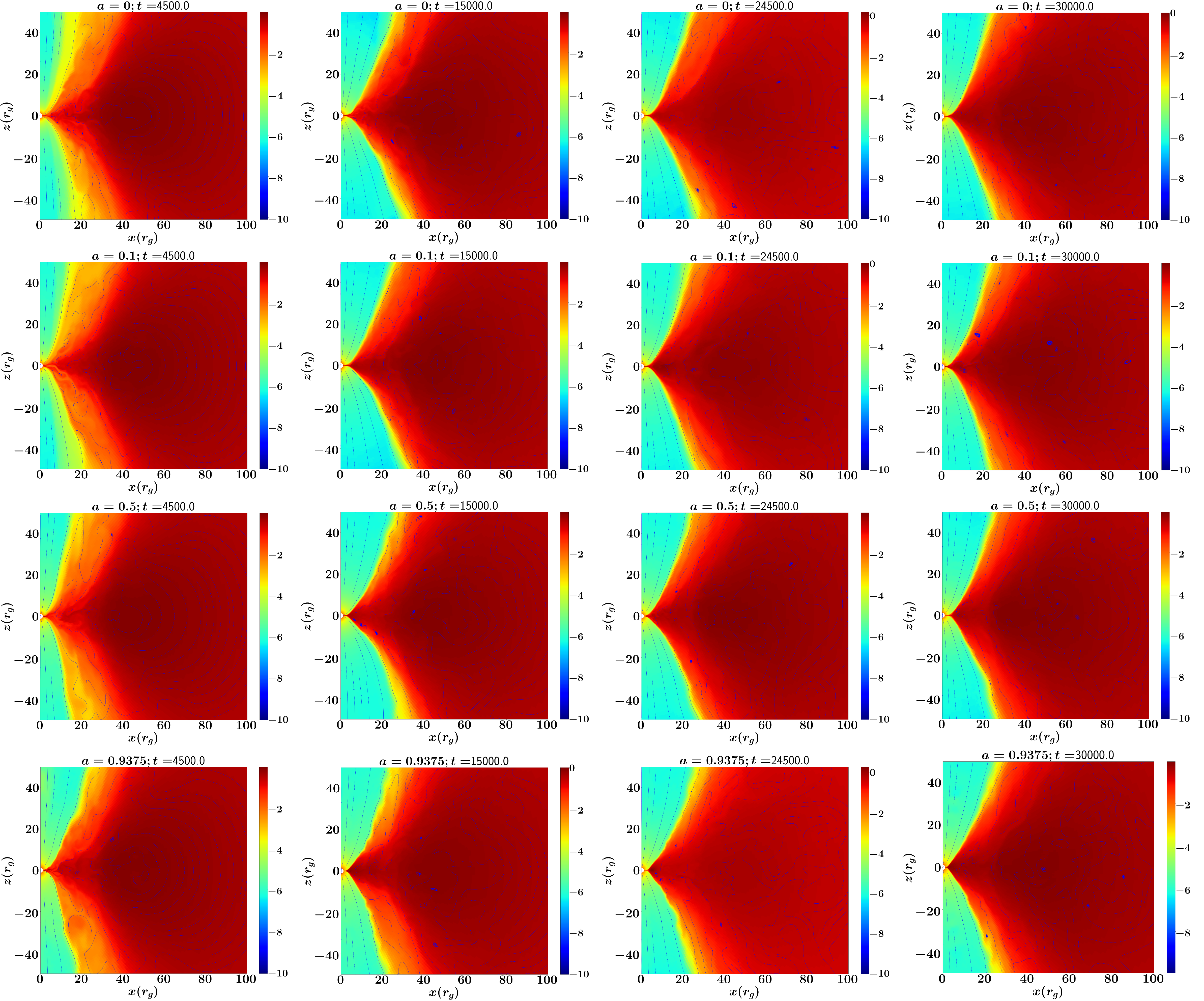}
     \caption{Logarithmic density contours at various time for SANE evolution with magnetic field streamlines for different spins of the black hole. For movie of density contour time evolution, follow: https://youtu.be/6SXBTk5ePes}
     \label{sane}
\end{figure*}

\begin{figure*}
\centering
\includegraphics[width=\textwidth]{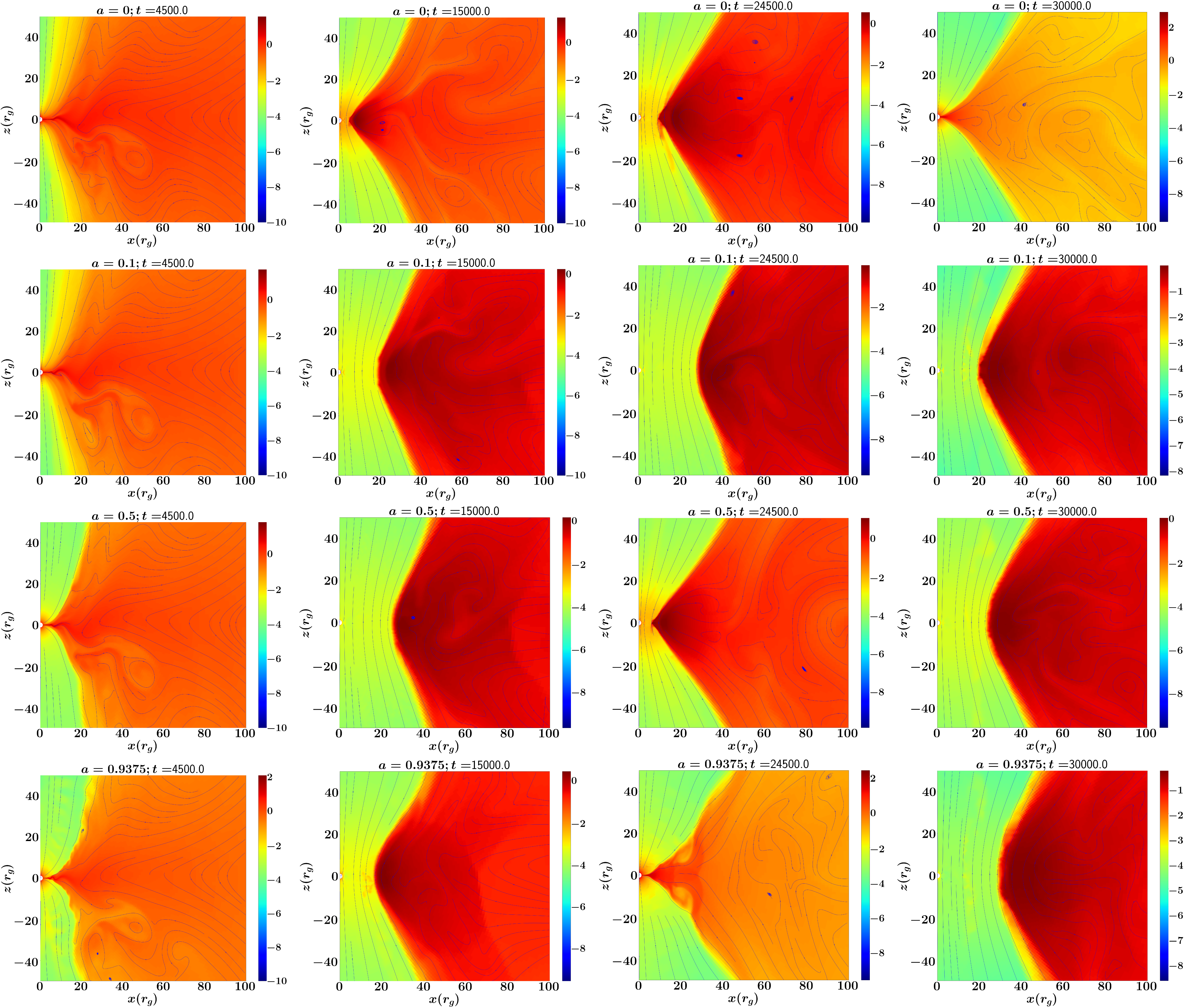}
     \caption{Logarithmic density contours at various time for MAD evolution with magnetic field streamlines for different spins of the black hole. For movie of density contour time evolution, follow: https://youtu.be/PCy37oTpykk}
     \label{mad}
\end{figure*}

\section{Simulation set-up} \label{sec:style}

We have used the GRMHD code BHAC (Black Hole Accretion Code, \citealt{bhac}) to initiate and evolve a magnetised accretion flow around a black hole. The Fishbone Moncrief (FM) torus solution \citep{fm} has been used to initiate matter density around the black hole. The code does not evolve the background spacetime geometry itself due to the low density of the flow and the reasonably short time evolution during which substantial change in the spacetime parameters is not expected. 

The equations solved by the code are:
\begin{equation}
\begin{aligned}
    \nabla_\mu(\rho u^\mu)&=0,\\
    \nabla_\mu T^{\mu\nu}&=0,\\
    \nabla_{\mu}\hspace{0.01in}^*F^{\mu\nu}&=0,   
\end{aligned}
\end{equation}
where $u^{\mu}$ is the four-velocity, $T^{\mu\nu}$ is the stress-energy tensor and $^{*}F^{\mu\nu}$ is the dual Faraday tensor. Here, $\mu$ and $\nu$ are spacetime indices such as $t,r,\theta,\phi$.

The code employs modified Kerr-Schild coordinates ($s,v$) to evolve the MHD equations. They convert a uniform Kerr-Schild ($r,\theta$) grid by the transformation: $r=e^s$ and $\theta=v+0.5h\sin2v$, where we have chosen $h=0.35$.
These coordinates concentrate resolution near the mid plane and close to the black hole, where most of the matter and magnetic fields reside. 

The boundary conditions used are as follows:

1. In the radial direction, the gradient of the primitive variables is set to zero at the boundaries of the simulation domain, i.e., at $r=1.22r_g$ and $r=2500r_g$, where $r_g=GM/c^2$ is the gravitational radius, $M$ is the mass of the black hole, $G$ is the Newton's gravitation constant, $c$ is the speed of light.

2. In the polar direction, hard boundary conditions are implemented. The flux through the poles of the domain ($\theta=0$ and $\pi$) is set to zero.

We use the equation of state as, $p=k\rho^\gamma$, where $\gamma = 5/3$, and $k$ is the polytropic constant. Here $\rho$ and $p$ are the density and pressure respectively.

We adopt geometric units, i.e., $GM=c=1$, hence $r_{g}=GM/c^2=1$ and the light crossing time, $t_g=GM/c^3=1$, in our simulations. We have carried out 2.5-dimensional simulations, by exploiting the axisymmetry of the system. The simulations have been run at a resolution of $384\times192\times1$. All simulations are evolved to $3\times10^4$ timesteps.
To study the effects of black hole spin on the outflow power, we consider four black hole spins in our simulations, $a=a_*/M=$ 0.0, 0.1, 0.5, 0.9375.

We explore two magnetic field evolution formalisms, namely SANE (standard and normal evolution) and MAD (magnetically arrested disk). The magnetic vector potential for these are given by \citep{fluxerr}:
\begin{enumerate}
    \item SANE: $A_{\phi}=\max(\rho/\rho_0-0.2,0)$,
    \item MAD: $A_\phi=\exp(-r/400)(r/r_{\mathrm{in}})^3\sin^3\theta\max(\rho/\rho_{0}-0.01,0)$,
\end{enumerate}
where $\rho_0$ is the maximum density in the initial torus, set at $r=41$, while $r_{\mathrm{in}}=20$ is the inner edge of the FM torus.
The initial field strength is set up by defining the initial plasma-$\beta$ parameter to be $100$ (plasma-$\beta=p_{\mathrm{gas}}/p_{\mathrm{mag}}=p_{\mathrm{gas}}/(b^2/2)$, where $p_{\mathrm{gas}}$ is the gas pressure and $b^2=b^\mu b_\mu$ is the norm of the four-magnetic field).
These simulations allow us to analyse magnetised accretion flows around black holes with different spins in separate magnetic field regimes, their implications on the outflow power and the overall evolution of the dynamics of the system.

Build up of large magnetic fields in a given volume leads to numerical errors in GRMHD codes \citep{ress}. To avoid this, we have set the maximum value of $b^2/\rho$ to be 100. Whenever this limit is exceeded, matter density is injected in the coordinate frame.
\begin{figure*}
\centering
\subfloat[SANE]{\includegraphics[width=0.49\textwidth]{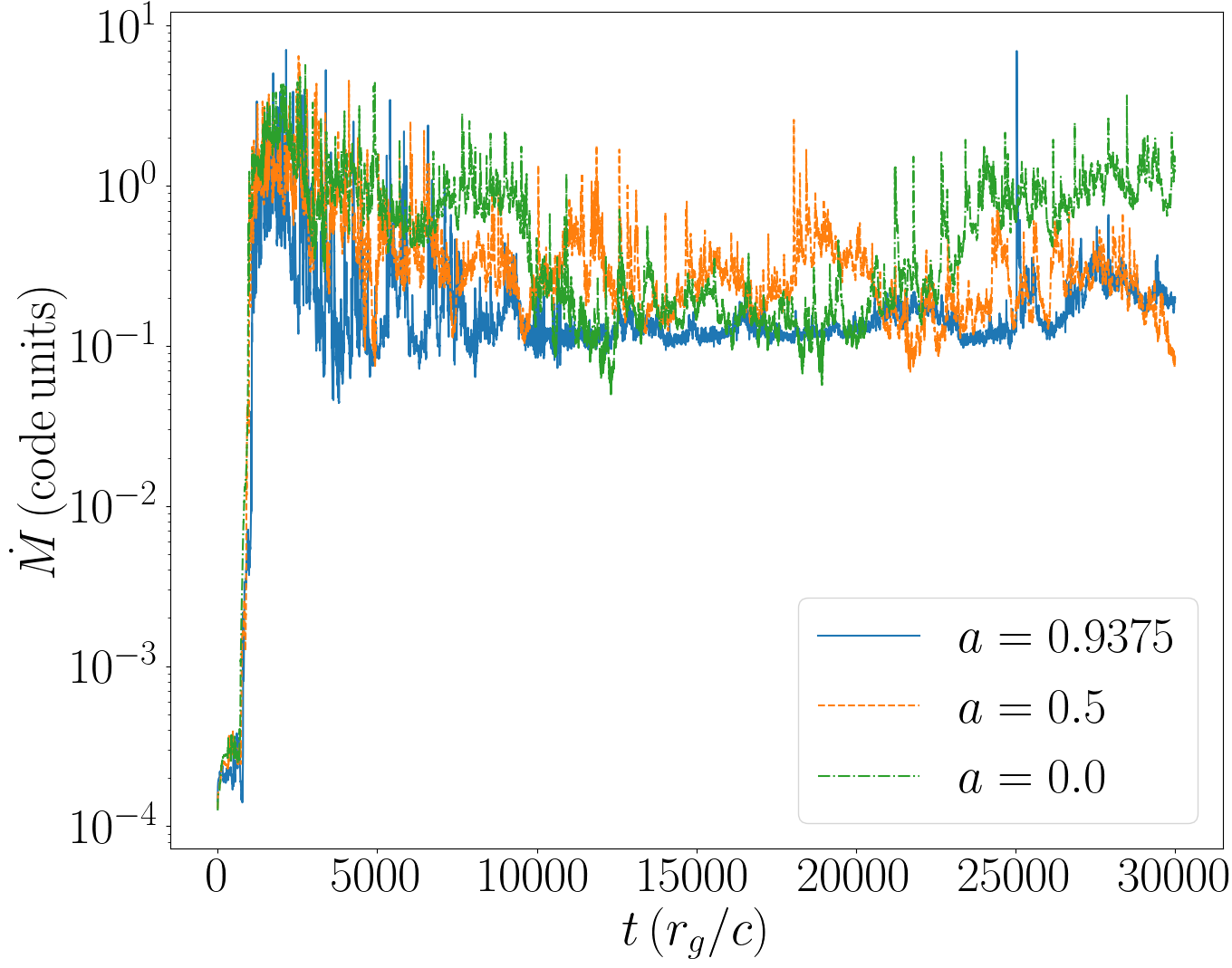}}
\subfloat[SANE]{\includegraphics[width=0.48\textwidth]{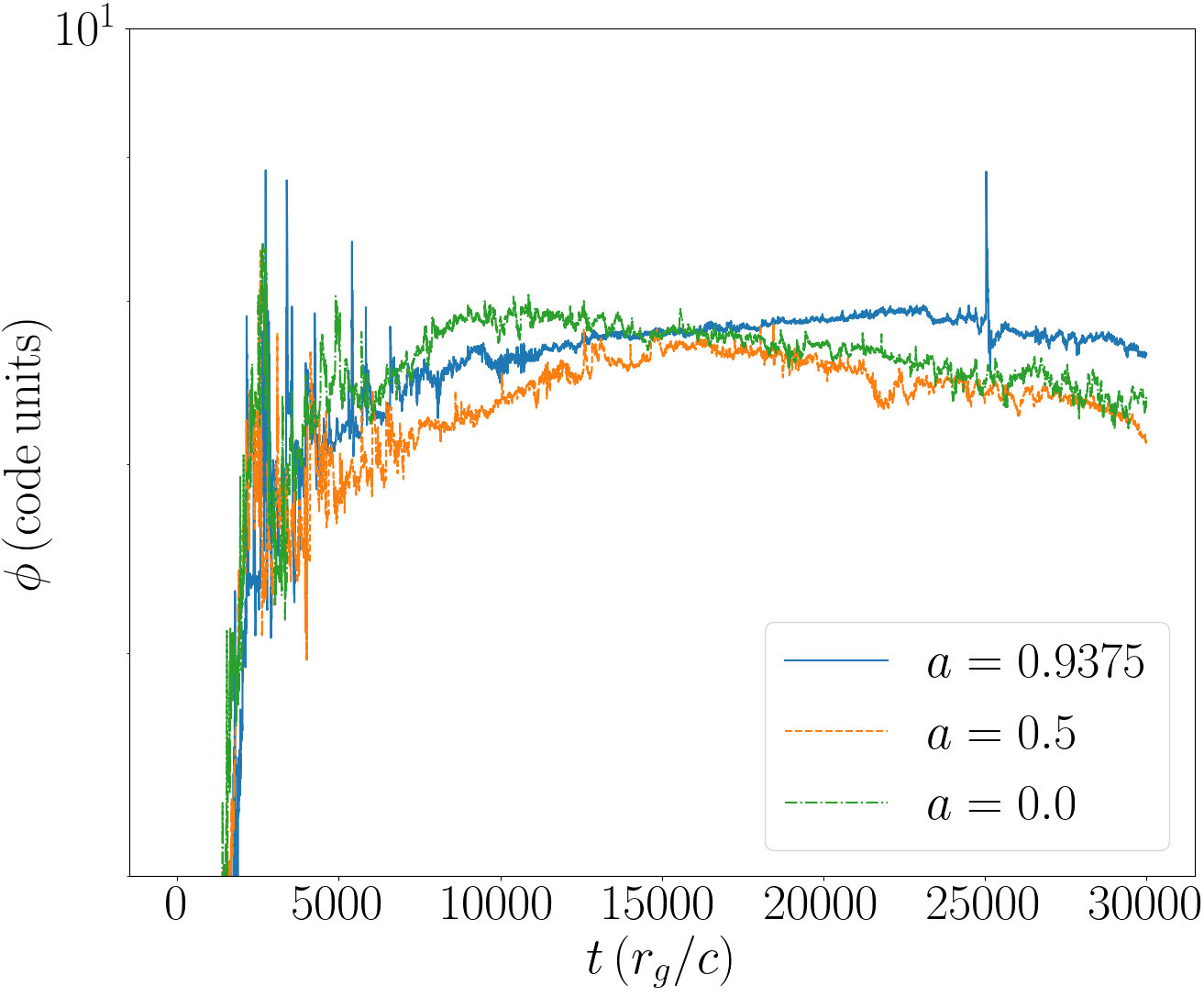}}

\subfloat[MAD]{\includegraphics[width=0.49\textwidth]{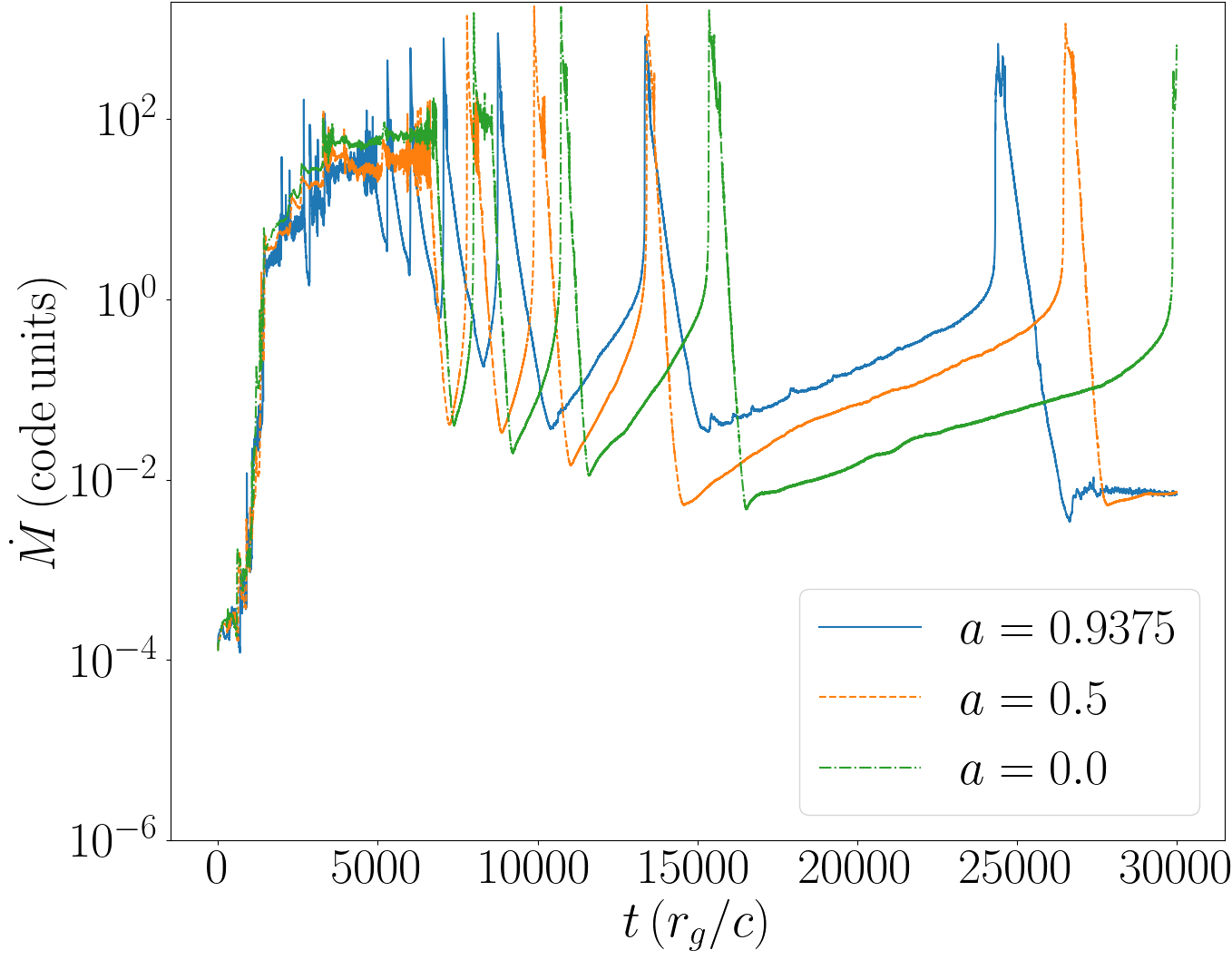}}
%\hspace{0.1in}
\subfloat[MAD]{\includegraphics[width=0.49\textwidth]{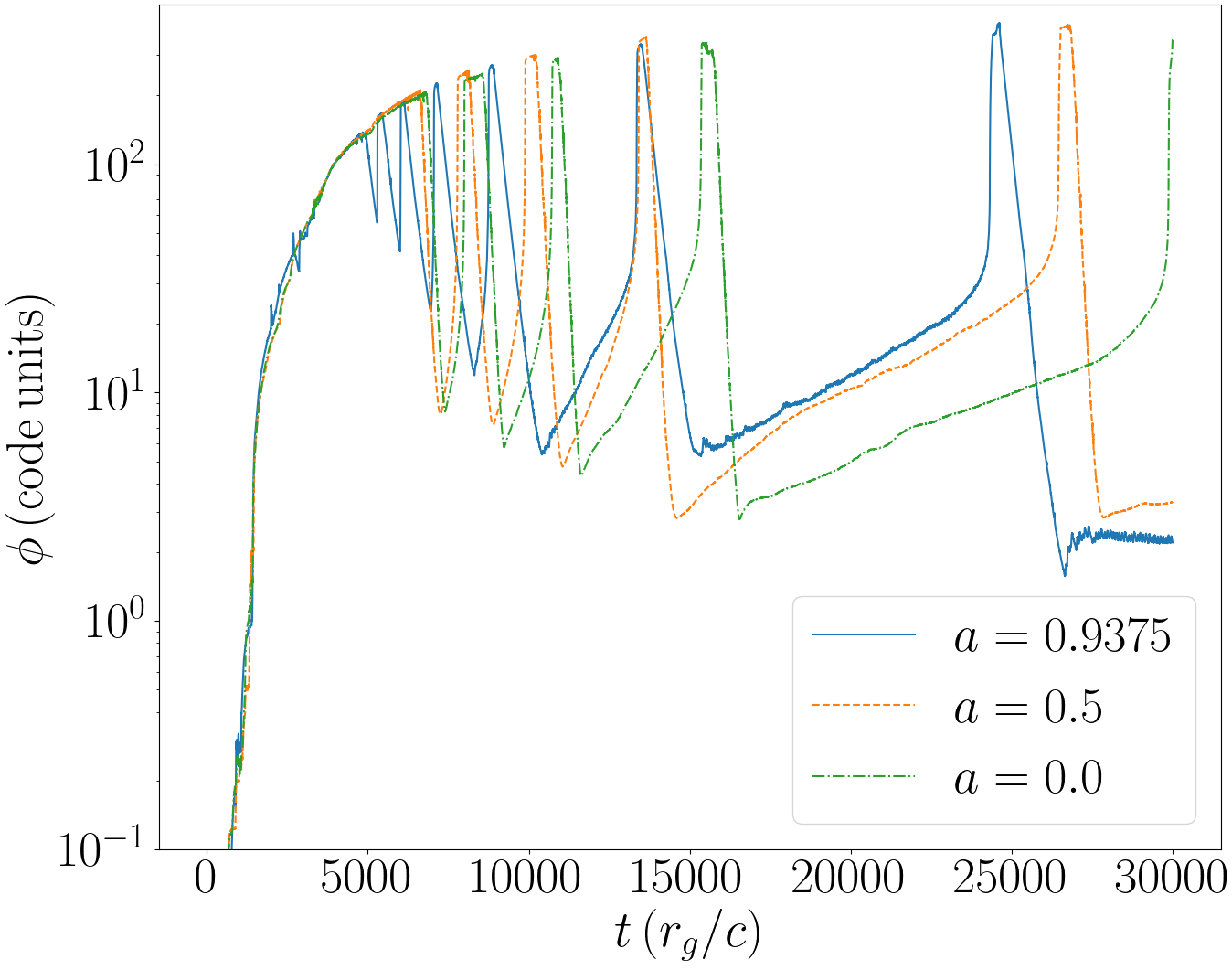}}
     \caption{The time evolutions of accretion rate and magnetic flux measured at the black hole event horizon for SANE and MAD simulations.}
     \label{t}
 \end{figure*}
\section{Simulation results and ULX}

The magnetic fields in SANE systems evolve slower than their MAD counterparts. This leads to different dynamics which in-turn affect the energetics of the system. This behaviour is evident in Figs. \ref{sane} and \ref{mad}, with MAD simulations showing more erratic behaviour than SANE. The magnetic field in MAD simulations builds up near the black hole, forming a barrier and throwing the matter outwards. Since we are working in the ideal MHD regime, the matter expelled in these eruption events also takes away some magnetic field with it due to flux freezing. This in turn leads to a relatively reduced magnetic field strength near the black hole, thus allowing the matter to accrete again onto the black hole and eventually again the fields build up. This cycle repeats throughout the evolution of the system, irrespective of the black hole spin. This effect is exaggerated in 2-dimensional simulations, as the accretion flow is confined to only one azimuthal plane.

On the other hand, in SANE simulations such eruptions are not seen for any of the black hole spins considered. This is due to the slower evolution of magnetic field in these systems. Field strength increases relatively slowly near the black hole and the matter remains in quasi-equilibrium with the magnetic field, thus not allowing any eruption event to occur. This is why we do not see any appreciable change in the density contours for SANE systems (Fig. \ref{sane}), as compared to MAD (Fig. \ref{mad}).

The temporal variations of the horizon accretion rate ($\Dot{M}$) and magnetic flux ($\phi$) for all the spin parameters ($a$) considered are shown in Fig. \ref{t}. It is evident that $\Dot{M}$ and $\phi$ reach a steady value in SANE simulations for all $a$ values. In MAD simulations, however, continuous increase and decrease in $\Dot{M}$ and $\phi$ are observed. This is because of the flux eruption events present in MAD systems mentioned above. Notice that the peaks in the $\Dot{M}$ and $\phi$ profiles occur at the exact same location. This is because we are working in the ideal MHD regime, hence the magnetic flux is frozen with the matter, resulting in a decrease in the magnetic flux along with matter eruption. The peak values of the magnetic flux in MAD simulations are also higher than SANE for all $a$. This further shows the magnetically dominated nature of MAD simulations.

To study the dynamics of the accretion flow, we consider the following quantities \citep{rn2022}:
\begin{enumerate}
    \item $\Dot{M}(r)=-\int\sqrt{-g}\rho u^r\mathrm{d}\theta \mathrm{d}\phi$.
    \item Energy flux: $\Dot{E}(r)=\int \sqrt{-g}T^r_t\mathrm{d}\theta \mathrm{d}\phi$, where $T^{\mu}_{\nu}=(\rho+p+u_g+b^2)u^{\mu}u_{\nu}+(p+b^2/2)\delta^\mu_\nu-b^{\mu}b_{\nu}$ is the stress-energy tensor.
    \item Outflow power: $P_c(r)=\Dot{M}-\Dot{E}$.
\end{enumerate}
Here, $\rho$ is the disk density, $p$ is the pressure of the flow, 
$\gamma$ is the adiabatic constant, $u_g=p/(\gamma+1)$ is the internal energy of the fluid, $u^{\mu}$ and $b^{\mu}$ are the four-velocity and four-magnetic field respectively, and $b^2=b^{\mu}b_{\mu}$. As evident from the above equations, energy flux is a quantity integrated over area of the domain under consideration. It is effectively the total power.
The sign of $\Dot{M}$ and $\Dot{E}$ is chosen such that mass and energy flow into the black hole is positive. The energy flux is defined with respect to an observer at infinity.

To study the radial variations of various quantities in the disk, we calculate their density averaged value over one disk scale height ($h$) which is calculated as \citep{codeco},
\begin{equation}
    \frac{h}{r}=\frac{\int\sqrt{-g}\rho |\theta-\pi/2|\mathrm{d}\theta \mathrm{d}\phi}{\int\sqrt{-g}\rho \mathrm{d}\theta \mathrm{d}\phi}.
\end{equation}
The results are then time-averaged over the last 15000 time-steps for all the simulations. For a quantity `$q$', its disk-average is thus given by \citep{mck12}

\begin{equation}
    <q>_{\mathrm{disk}}=\left.\frac{\int q\sqrt{-g}\rho  \mathrm{d}\theta \mathrm{d}\phi}{\int\sqrt{-g}\rho \mathrm{d}\theta \mathrm{d}\phi}\right\rvert_{\textrm{scale height}}.
\end{equation}

\subsection{Accretion rate profiles}

\begin{figure}
\centering
 \subfloat[SANE]{\includegraphics[width=0.48\textwidth]{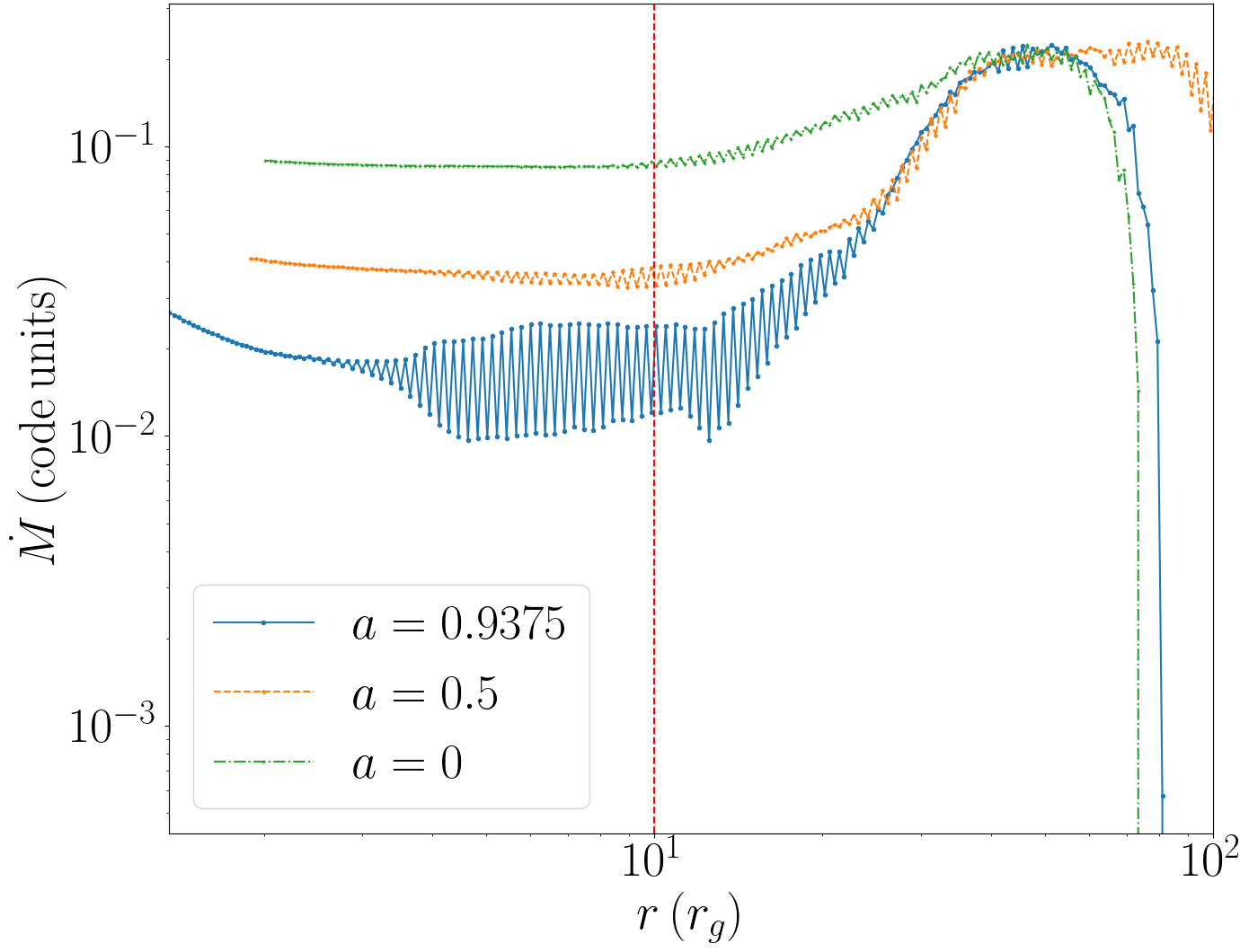}}

 \subfloat[MAD]{\includegraphics[width=0.48\textwidth]{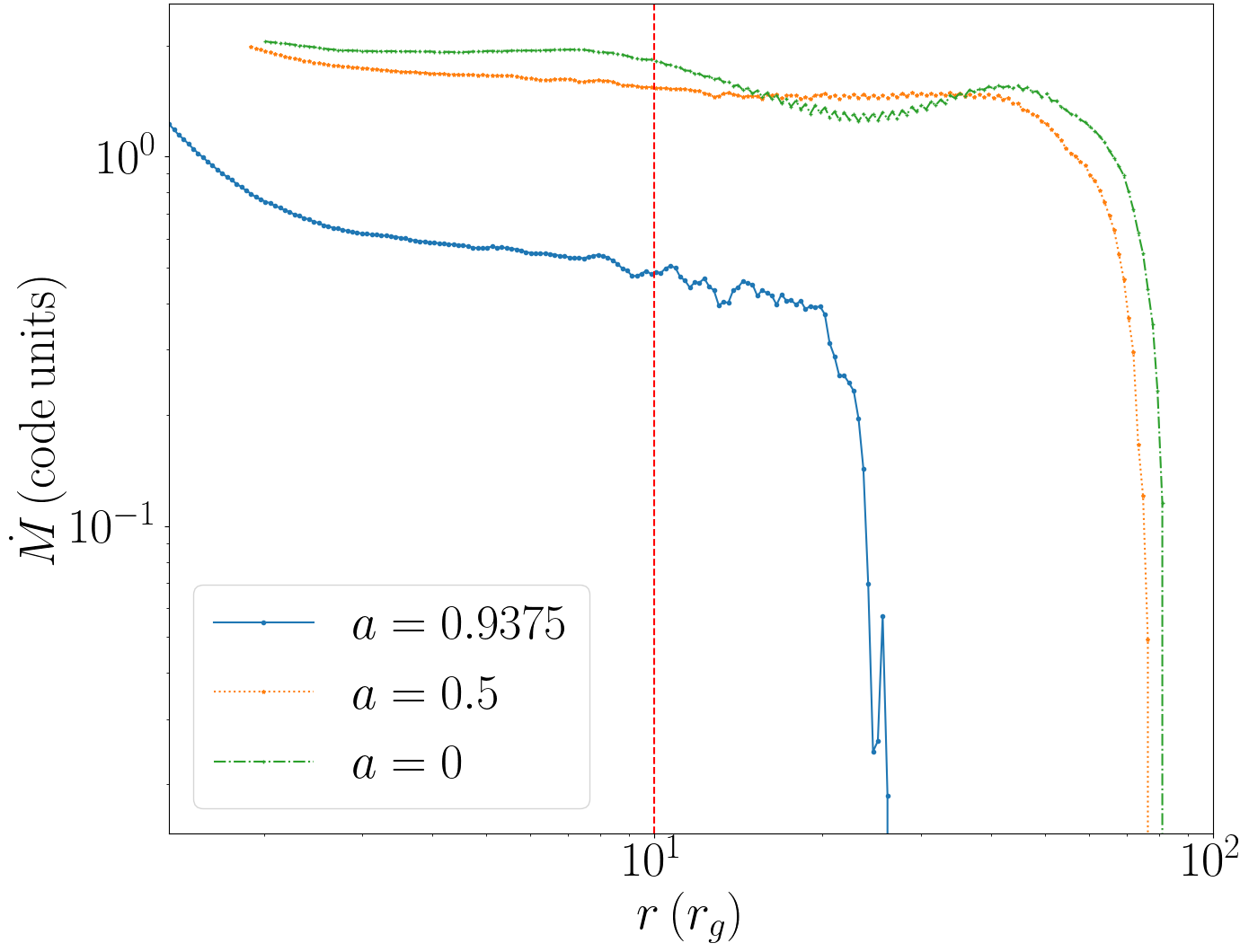}}
     \caption{Accretion rate profiles for SANE and MAD simulations.}
     \label{acc}
 \end{figure}

The accretion flow at a given radius is subject to inflow and outflow. As the simulation evolves, the flow reaches a state of inflow-outflow equilibrium out to a certain radius \citep{fluxerr}. We refer to this radius as the steady flow radius ($r_{\mathrm{eq}}$). Depending on the initial magnetic vector potential and time of evolution, $r_{\mathrm{eq}}$ changes.

\begin{figure*}
\centering
 \subfloat[SANE]{\includegraphics[width=0.49\textwidth]{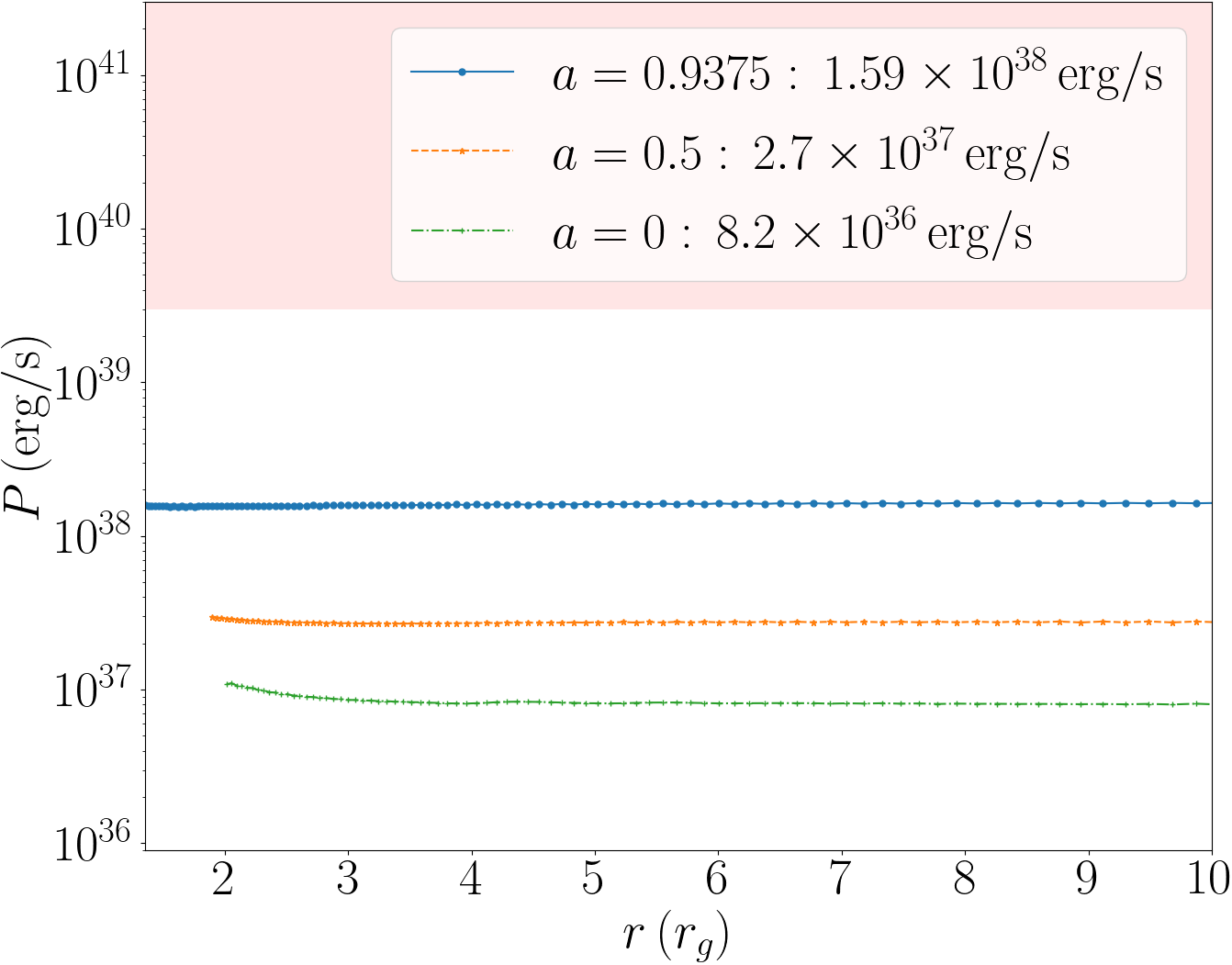}}
 \subfloat[MAD]{\includegraphics[width=0.49\textwidth]{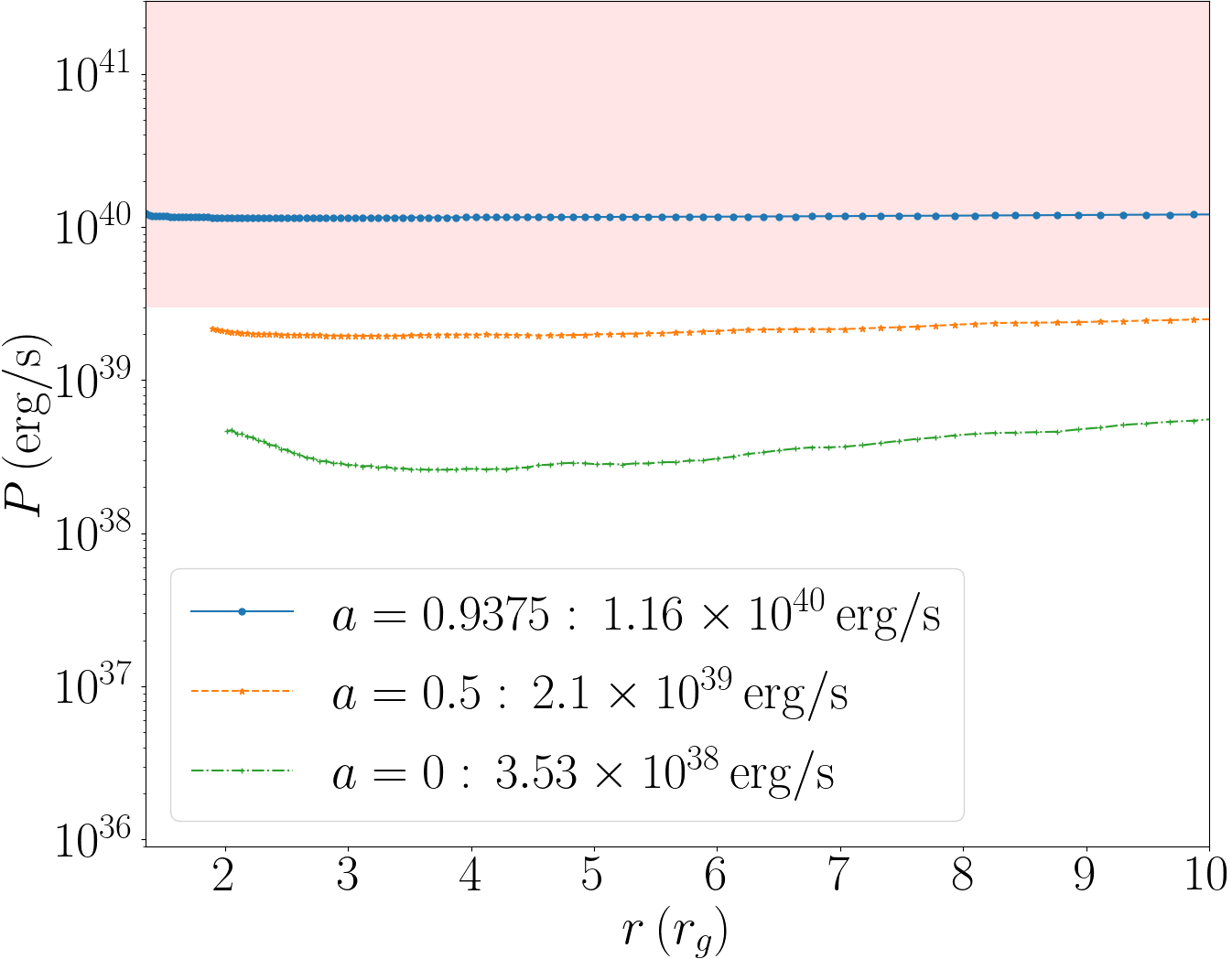}}
     \caption{Outflow power profiles for SANE and MAD simulations. Shaded region indicates the observed ULX luminosity range. The average outflow power from the inflow-outflow equilibrium region is mentioned in the legend.}
     \label{power}
 \end{figure*}
\subsection{Outflow power profiles}

Fig. \ref{acc} shows the obtained time-averaged accretion rate profiles of SANE and MAD simulations for different $a$. As evident, both SANE and MAD simulations reach the inflow-outflow equilibrium state for all $a$. Although for some $a$, the accretion rate profile is flat till larger radii before eventually dropping off, we have chosen $r_{\mathrm{eq}}=10$ because this is the optimum radius so that by this all the simulations reach inflow-outflow equilibrium. This also facilitates comparison among all the simulations considered. These $r_{\mathrm{eq}}$ and $\Dot{M}(r_{\mathrm{eq}})$ will be used in the following section to quantify the outflow power.

\subsection{Outflow power calculation}

The time evolution of various system parameters is governed by the stress energy tensor of the system. 
The net outflow power of the system depends on $\Dot{M}$ and $\Dot{E}$ and is defined by (in code units)
\begin{equation}
    P_c(r)=\Dot{M}(r)-\Dot{E}(r) \label{pow}.
\end{equation}

To compare our results to observations, we calculate the dimension-full outflow power using equation (\ref{pow}). For this purpose, the dimension-full factor $\Dot{M}_s c^2$ is to be multiplied to $P_c(r)$, where $\Dot{M}_s$ is the scale accretion rate, given by $\Dot{M}_s=\Dot{M}_{\mathrm{phy}}/\Dot{M}_{\mathrm{code}}(r_{\mathrm{eq}})$, with $\Dot{M}_{\mathrm{phy}}$ being the physical accretion rate  and $\Dot{M}_{\mathrm{code}}(r_{\mathrm{eq}})$ is the accretion rate in code units at $r=r_{\mathrm{eq}}$.

As evident from Fig. \ref{acc}, the MAD simulations have higher accretion rates than SANE for all $a$. To obtain $\Dot{M}_s$, we scale the accretion rate of the MAD flow with $a=0.9375$ such that it has $\Dot{M}_{\mathrm{phy}}=0.05\Dot{M}_{\mathrm{Edd}}$ at $r=10$, where $\Dot{M}_{\mathrm{Edd}}$ is the Eddington accretion rate, given by $\Dot{M}_{\mathrm{Edd}}=1.39\times10^{18}(M/M_{\odot})$ gm/s and $M$ is chosen to be $20M_{\odot}$ throughout following ULX19. The corresponding $\Dot{M}_s$ is used to calculate the outflow power for the other spins as well for both SANE and MAD systems.
The dimension-full outflow power is thus defined by
\begin{equation}
    P(r)=\left(\frac{\Dot{M}(r)-\Dot{E}(r)}{\Dot{M}(r_{\mathrm{eq}})}\right)\Dot{M}_{\mathrm{phy}}c^2\hspace{0.05in}\mathrm{erg/s}.
    \label{dpower}
\end{equation}
Here, $P(r)$ is calculated by summing over contributions from entire range of $\theta$ at each $r$.
The obtained power profiles are shown in Fig. \ref{power}. These profiles show that the outflow power increases with the black hole spin, which has also been obtained by \cite{rn2022}. This confirms that faster spinning black holes are capable of producing more powerful outflows, possibly due to an increase in contribution to the outflow power by the Blandford-Znajek mechanism \citep{bz}. Moreover, the outflow profiles also show that MAD systems produce more powerful outflows than SANE systems. This can be attributed to the fact that MAD systems are more efficient to extract energy than SANE systems. 

As evident from Fig. \ref{power}, only the MAD system with $a=0.9375$ has outflow power in the observed ULX luminosity range. This shows that hard state ULXs can be interpreted as magnetically arrested advective accretion systems around spinning stellar mass black holes.

\cite{rn2022} showed that the outflow efficiency (and hence outflow power) increases monotonically with the spin of the black hole in accordance with the Blandford-Znajek power prediction. From Fig. \ref{power}(b), it is clear that black hole spins higher than $a=0.5$ can lie in the shaded region of observed ULX luminosities. This further shows that considering ULXs to be stellar mass black hole sources, to achieve such high luminosities, the black hole spin must be on the higher side.

\subsection{Magnetic field profiles}

\begin{figure}
 \centering
 \subfloat[SANE]{\includegraphics[width=0.48\textwidth]{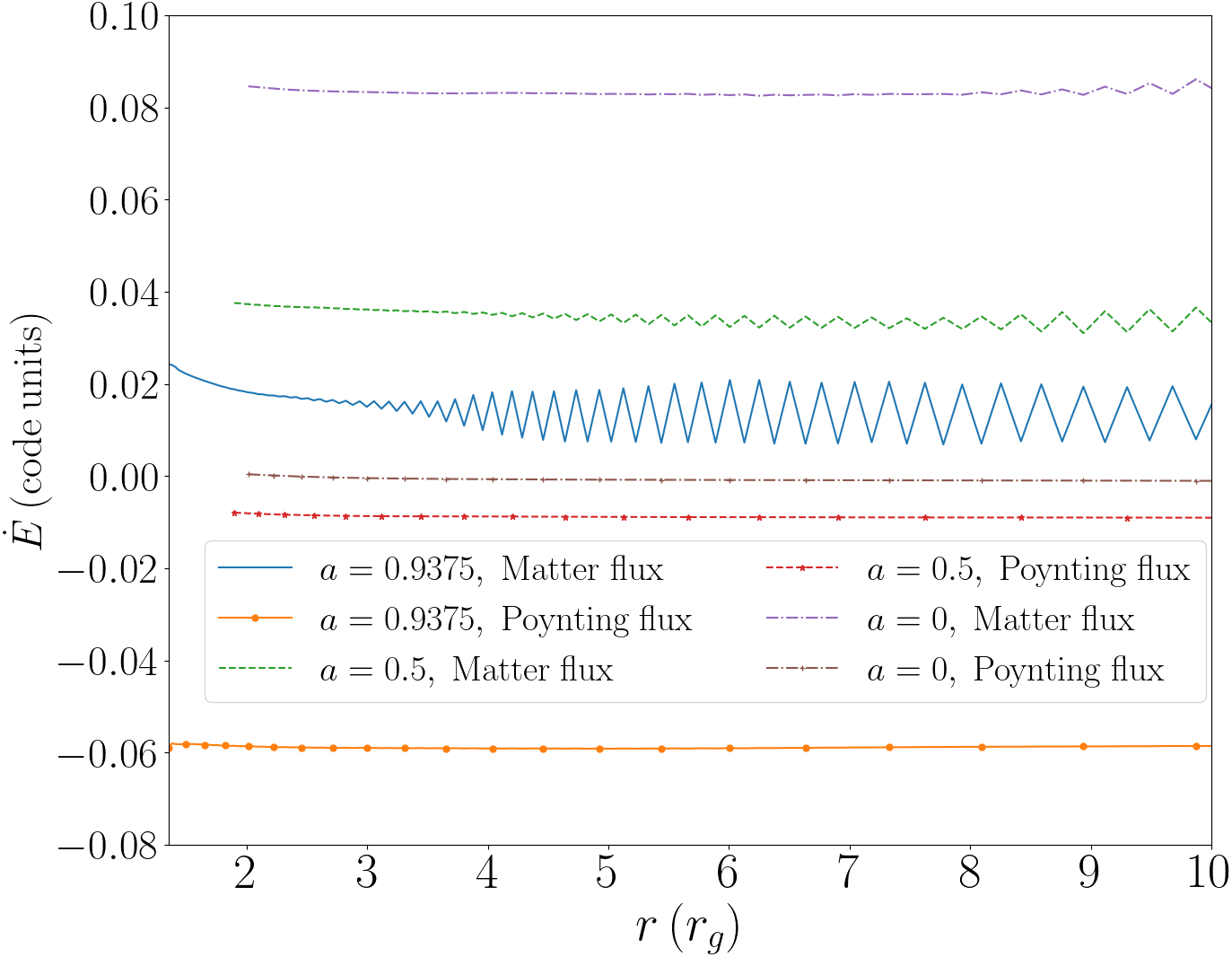}}

\subfloat[MAD]{\includegraphics[width=0.48\textwidth]{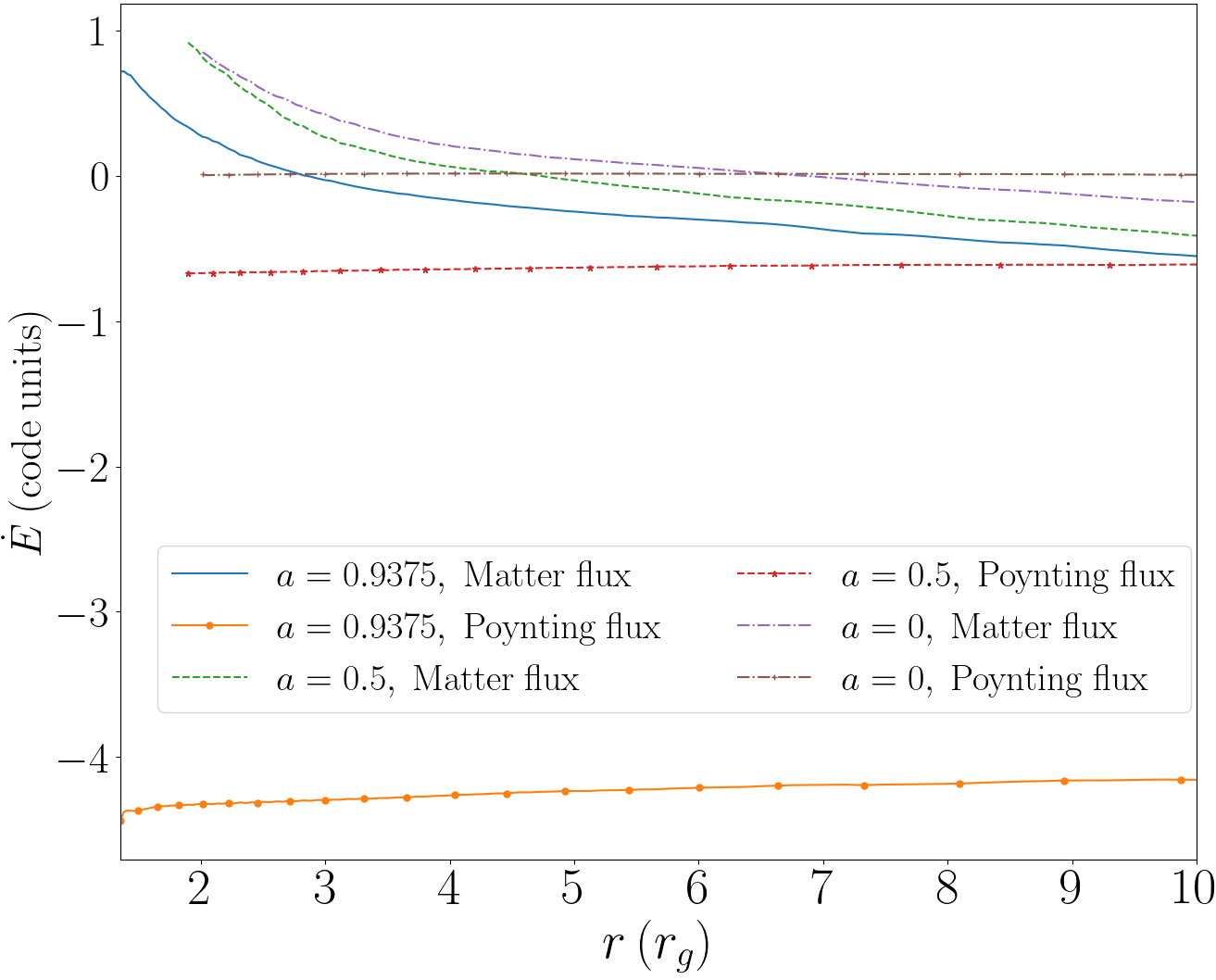}}
     \caption{Comparison of the magnetic energy flux with the matter energy flux.}
     \label{flux}
 \end{figure}

In Fig. \ref{flux}, we show the time averaged magnetic and matter part of the total flux, which are respectively given as, $\Dot{E}_{\mathrm{mag}}=\int\sqrt{-g}(b^2u^{r}u_{t}-b^{r}b_{t})\mathrm{d}\theta \mathrm{d}\phi$ and $\Dot{E}_{\mathrm{matter}}=\int\sqrt{-g}(\rho+p+u_g)u^{r}u_{t} \mathrm{d}\theta \mathrm{d}\phi$. It is evident from Fig. \ref{flux} that the magnitude of magnetic energy flux is several times higher than the matter energy flux for high-spinning black holes for both MAD and SANE systems.
For intermediate to low spinning black holes, however, the matter energy flux is higher than the the magnitude of magnetic energy flux.
Note that negative $\dot{E}$ implies that the total energy flow is away from the black hole.
This illustrates the magnetically dominated nature of the accretion flow around high-spinning black holes, while lower-spinning black holes have matter dominated energy flux. Although the matter energy flux for the non-spinning black hole case is higher than the magnitude of the magnetic energy flux of the high-spinning black hole in our SANE simulations, the energy returned to infinity by the accretion flow, i.e. outflow, is the difference between the mass accretion energy and energy flux. This confirms that high spinning black holes are more prone to produce outflows and jets with efficiency, $P_c/\dot{M}$, greater than unity \citep{rn2011}. As evident from Fig. \ref{acc}, non-spinning black holes also have the highest mass accretion rate. This leads to the net outflow power from these systems being the lowest. In MAD simulations, however, the magnetic energy flux is much higher compared to the matter energy flux of lower-spinning black holes.

This explains the high luminosities obtained for high spinning MAD systems. Due to the abundance of magnetic fields in this system, the accretion flow can tap the Poynting energy more efficiently.
As the energy flux is dominated by the magnetic component, it is important to analyse the magnetic field profiles in the accretion flow.

\begin{figure}
\centering
\subfloat[SANE]{\includegraphics[width=0.48\textwidth]{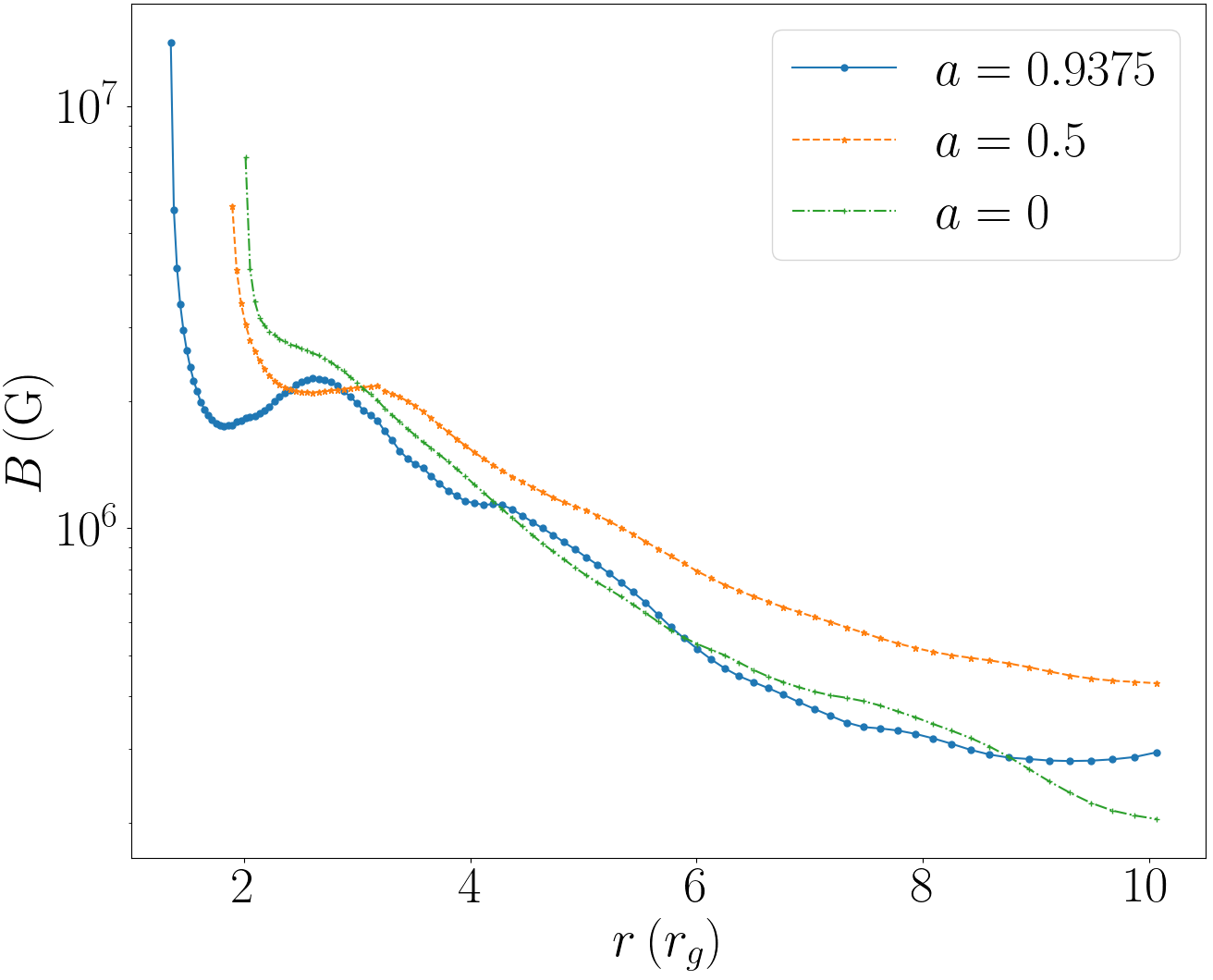}}

\subfloat[MAD]{\includegraphics[width=0.48\textwidth]{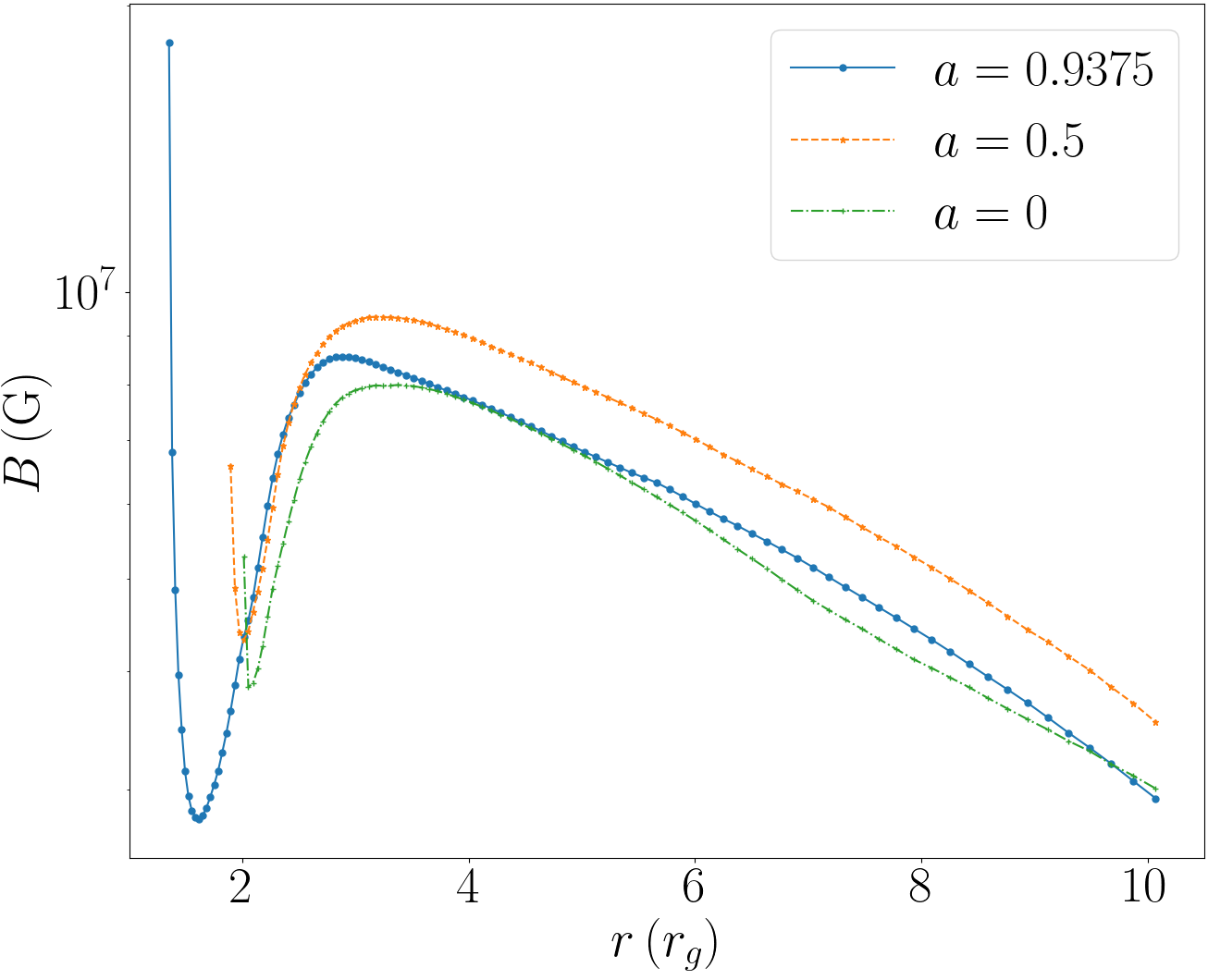}}
     \caption{Magnetic field profiles in the Eulerian frame.}
     \label{mag}
 \end{figure}

Fig. \ref{mag} shows the time averaged and disk averaged magnitude of the Eulerian magnetic field profiles which are given by the transformation \citep{bhac}:
\begin{equation}
    B^i=\alpha(b^iu^t-b^tu^i),
\end{equation}
where $\alpha$ is the lapse function, $b^i$ and $u^i$ are the spatial components of the four magnetic field and four velocity respectively, $b^t$ and $u^t$ are the temporal components of the four magnetic field and four velocity respectively.

As evident from Fig. \ref{mag}, the peak magnetic field magnitude for MAD with $a=0.9375$ is the highest among all the systems. This explains the higher outflow power obtained for this system. The field is maximum at the horizon due to the higher angular velocity of the black hole coiling (and thus accumulating) the field around the axis of the black hole. Away from the horizon, the field drops between the first and second peaks due to the reduced effect of black hole spin. All the MAD systems show a second peak in magnetic field at around $r=3$. This peak may be attributed to the resultant effect of the flux eruption events in MAD systems. The continuous back and forth of the accretion flow in the simulation leads to an equilibrium position for the magnetic fields (around $r=3$) below which the field reduces due to eruption events and above which also the field reduces due to increasing distance from the black hole. No such peak is seen in the SANE case due to the absence of flux eruption events.

The field strength in the MAD case with $a=0.9375$ is around $2\times10^7$ G. This is in accordance with the field strength obtained by ULX19, which is required for producing such high luminosities in hard state ULXs.

Although SANE with $a=0.9375$ also has the peak magnetic field magnitude of the order of $10^7$ G, it does have a low outflow power due to lower mass accretion rate (Fig. \ref{acc}a) and lower energy flux (Fig. \ref{flux}a). The MAD systems also have consistently higher magnetic fields than their SANE counterparts, which explains their higher outflow power.

\section{FSRQ/BL Lac dichotomy}

\begin{figure}
\includegraphics[width=0.48\textwidth]{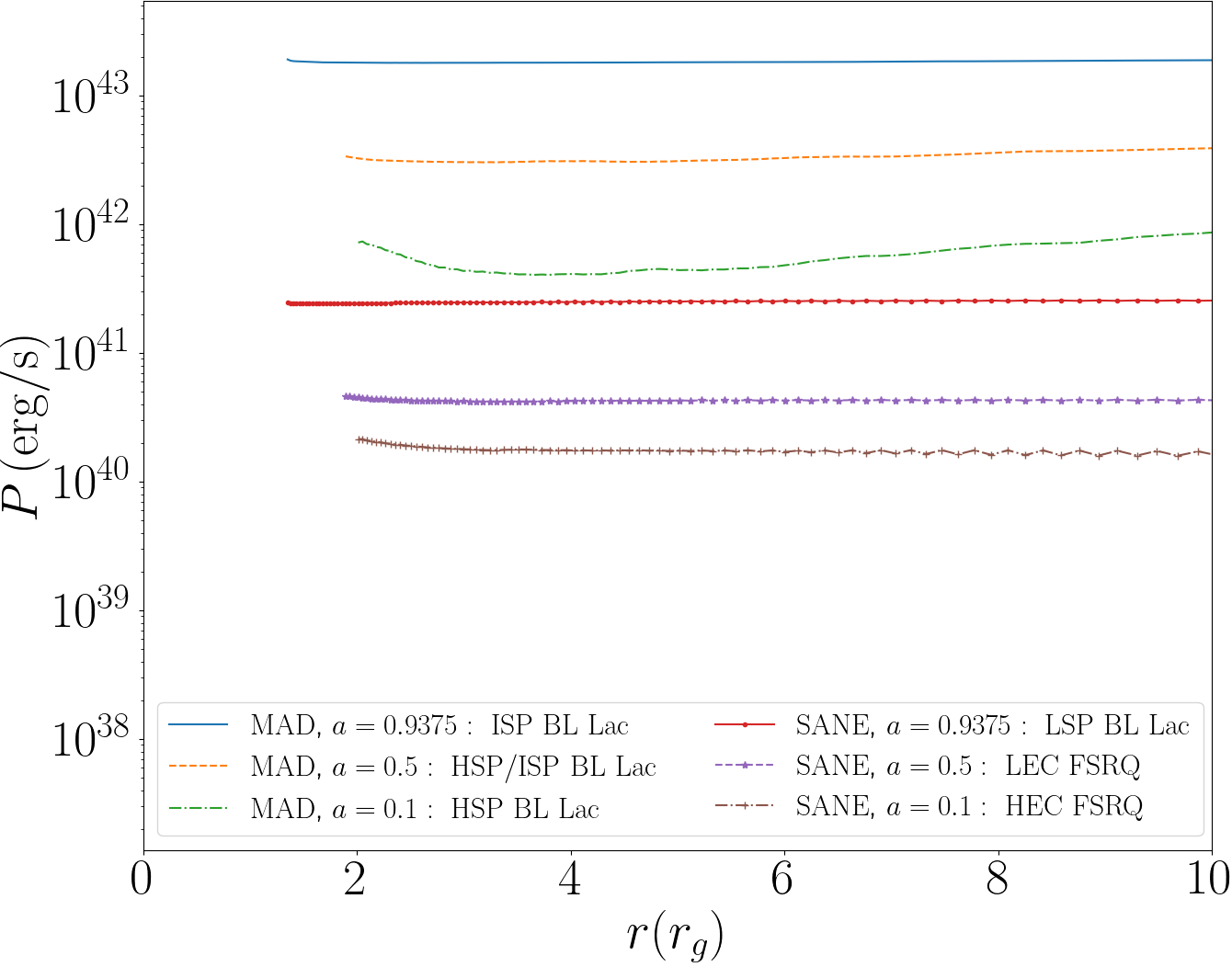}
     \caption{FSRQ and BL Lac power profiles.}
     \label{fsrqpo}
 \end{figure}

From our simulations, we have interpreted the FSRQ/BL Lac dichotomy as the result of their magnetic flux saturation leading to different accretion dynamics and thus dissimilar intrinsic outflow power.
Using the same setup as described in section 2, we have explored our simulated results for three different black hole spins, i.e. $a=0.1, 0.5, 0.9375$, by appropriately scaling the mass and accretion rate as described below.

As described in section 3.2, to compare our simulation results with observations, we need to restore the dimensions of the calculated outflow power. To do so, we set an appropriate $\Dot{M}_s$, which will be used to scale the accretion rate and outflow power for all the simulations considered for FSRQ/BL Lac.

BL Lac objects are widely considered to be advective accretion flows \citep{ct95,adaf} due to their non-thermal spectra \citep{mar03}. One of the most talked about sources is M87 which seems to be BL Lac \citep{jong15}. Via EHT observations and numerical simulations, the accretion flow around M87 has been well understood to be a MAD system \citep{tsv98,cyl23}. Following these results, we choose the $\Dot{M}_s$ for our analysis to be $5\times10^{-5}\Dot{M}_{\mathrm{Edd}}$ with $M=10^8 M_{\odot}$. This choice puts the accretion rate for all our MAD results in the advective regime and also scales the computed outflow power to be equivalent to the debeamed luminosities calculated by BL19. 

The dimension-full outflow powers are shown in Fig. \ref{fsrqpo}. As evident, MAD simulations have higher outflow power than SANE systems. We interpret the low spin SANE systems as high EC (HEC) FSRQs and the intermediate spin SANE systems as low EC (LEC) FSRQs. This is because the accretion rate scales inversely with spin (Fig. \ref{acc}). Lower accretion rate corresponds to less number of soft photons available for EC. This leads to less EC and hence higher intrinsic luminosity which is clearly evident from our simulation results. BL19 also showed that increasing EC fraction leads to lower intrinsic luminosities in FSRQ systems.

Comparing our results with the analysis of BL19, we also conclude that MAD systems with high spins ($a=0.9375$) correspond to ISP BL Lacs as they have the highest intrinsic luminosity out of all the simulations considered, and MAD system with $a=0.1$ corresponds to HSP BL Lacs. The intermediate spin case ($a=0.5$) can be interpreted as a combination of HSP and ISP BL Lacs.

BL19 also argued that the similarities in the observed $\gamma$-ray luminosities of LSP BL Lacs and FSRQs can be attributed to the fact that LSP BL Lacs have a substantial EC component along with the SSC component. This shows that LSP BL Lacs are quite similar to FSRQs. Following this line of reasoning, we have interpreted LSP BL Lacs as high spin SANE systems. 

We have thus shown by our simulations that the apparent dichotomy between FSRQs and BL Lacs is due to their intrinsic magnetic properties. The corresponding EC and synchrotron peak characteristics can be attributed to the spin of the central supermassive black hole.

\subsection{Magnetic field stresses}

\begin{figure}
\centering
 \subfloat[MAD]{\includegraphics[width=0.48\textwidth]{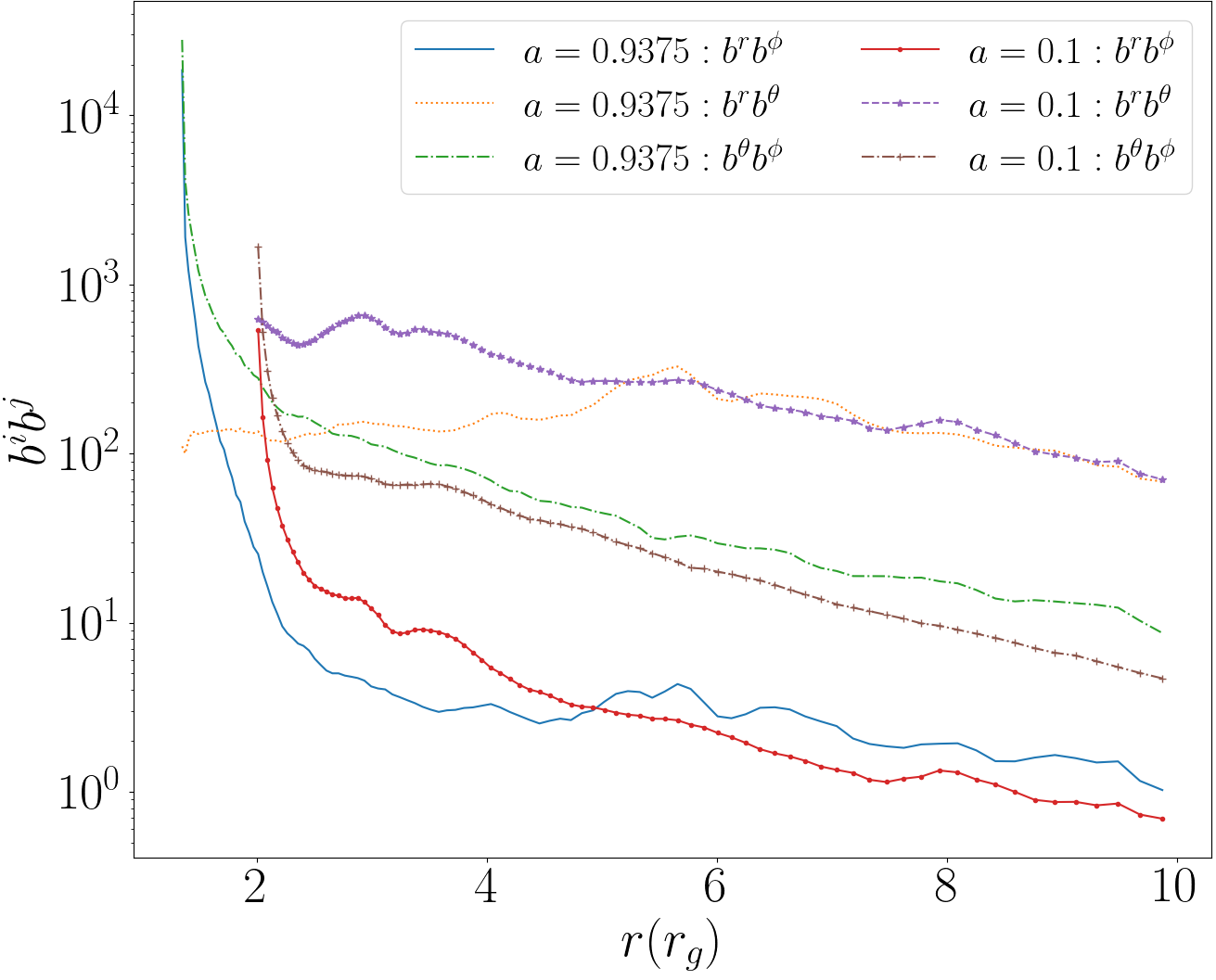}}

\subfloat[SANE]{\includegraphics[width=0.48\textwidth]{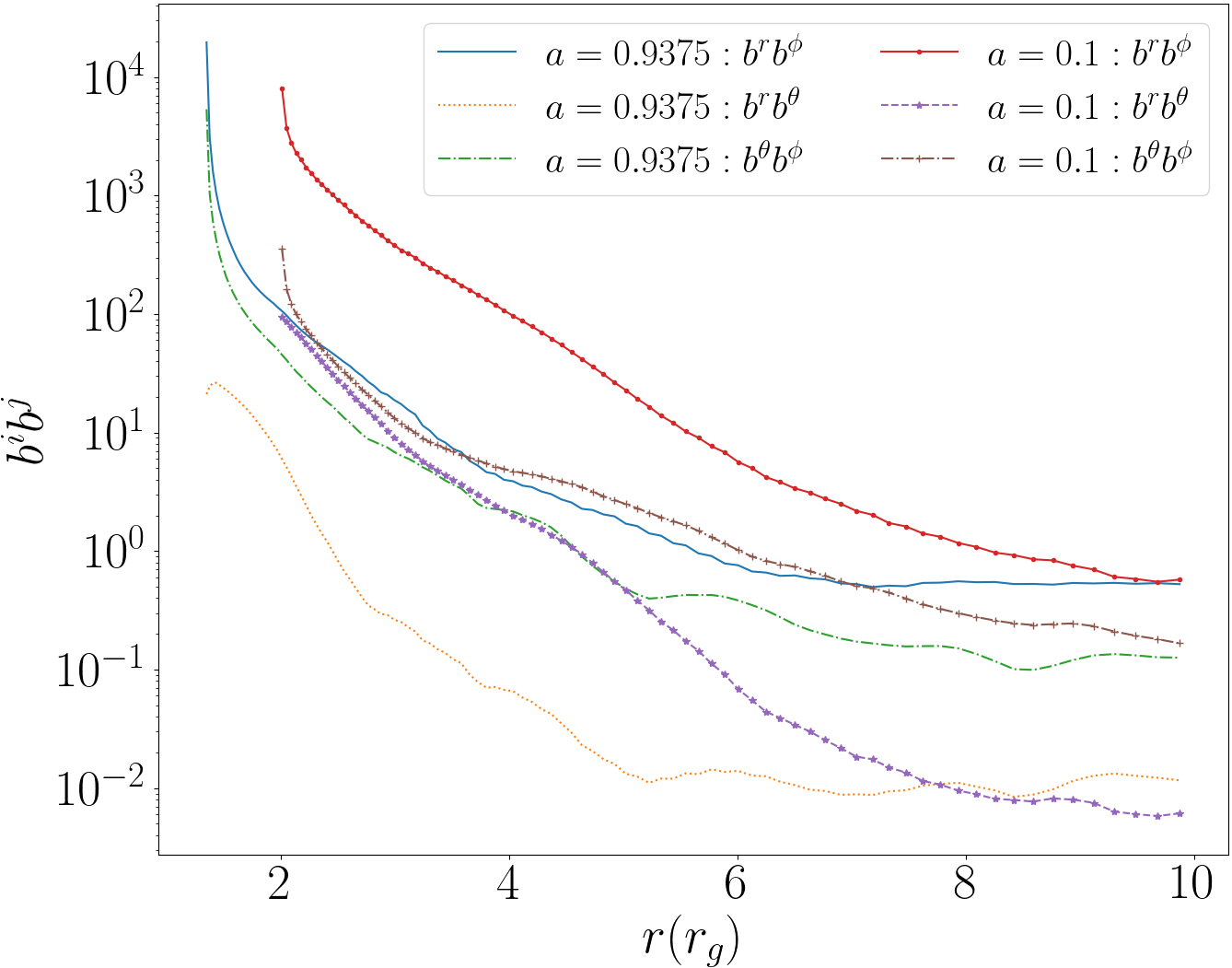}}
     \caption{MAD and SANE magnetic field stresses.}
     \label{stress}
 \end{figure}
 
Fig. \ref{stress} shows the four-magnetic field stresses ($b^ib^j$) for SANE and MAD simulations for high and low spin systems. Based on the conclusions made above, it is evident that BL Lac systems are mostly MAD and are dominated by the $b^{\theta}b^\phi$ component close to the black hole. This component contributes to jet launching from the ergosphere by the Blandford-Znajek (BZ) mechanism \citep{bz}. Comparing the low spin and high spin results, it is clear that $b^{\theta}b^\phi$ increases with black hole spin, thus leading to high jet power as predicted by the BZ mechanism. Outside the ergosphere, the $b^{r}b^\theta$ component dominates. This shows in Fig. \ref{stress}(a) that $b^{r}b^\theta$ will contribute to wind-like poloidal outflows from the accretion system. Thus BL Lacs are dominated by large scale poloidal stresses outside the ergosphere as also shown by BL19.

Fig. \ref{stress}(b), on the other hand, shows that FSRQs (which are understood to be SANE from the above discussion) are dominated by the $b^rb^\phi$ component throughout the disk. The $b^r b^\phi$ stress is a result of the rotation of the disk leading to an increase in $b^\phi$ due to coiling of the magnetic field lines due to flux freezing. This leads to disk winds dominated by toroidal magnetic fields \citep{indu}. As the plasma-$\beta>1$ in this case throughout (see \S 4.3 for details), it cannot be inferred with certainty whether this outflow will become a jet via the Blandford-Payne process \citep{bp}.

For the high spinning case, the $b^\theta b^\phi$ component becomes comparable to $b^rb^\phi$ due to the ergosphere effect of the black hole. This signifies that inside the ergosphere the outflow is generated due to both $b^\theta b^\phi$ and $b^r b^\phi$ components. Outside the ergosphere, however, the $b^r b^\phi$ stress becomes several times higher than the $b^\theta b^\phi$ stress, independent of $a$.
It is also evident from Figs. \ref{stress}(a) and (b) that $b^r b^\phi$ is also of similar magnitude in MAD and SANE systems. This shows similar amount of outflows driven by this magnetic stress component in both these systems. In MAD, however, the outflows due to $b^\theta b^\phi$ and $b^r b^\theta$ dominate. This further explains the much higher outflows in MADs compared to SANE.
MAD flows (BL Lacs) are thus dominated by large scale magnetic fields. In SANE flows (FSRQ), however, it is the plasma-$\beta$ which determines the contribution for large and small scale magnetic fields,  dictating the flow. Hence, the latter may be dominated by the magneto-rotational instability (MRI) process. This is similar to what was obtained by BL19 as well.

This shows that BL Lac (or MAD) systems are magnetically dominated, whereas FSRQs (or SANE) are disk dominated systems. Further validations of this assertion are provided in the following sections.

\subsection{Angular momentum}

\begin{figure}
\includegraphics[width=0.48\textwidth]{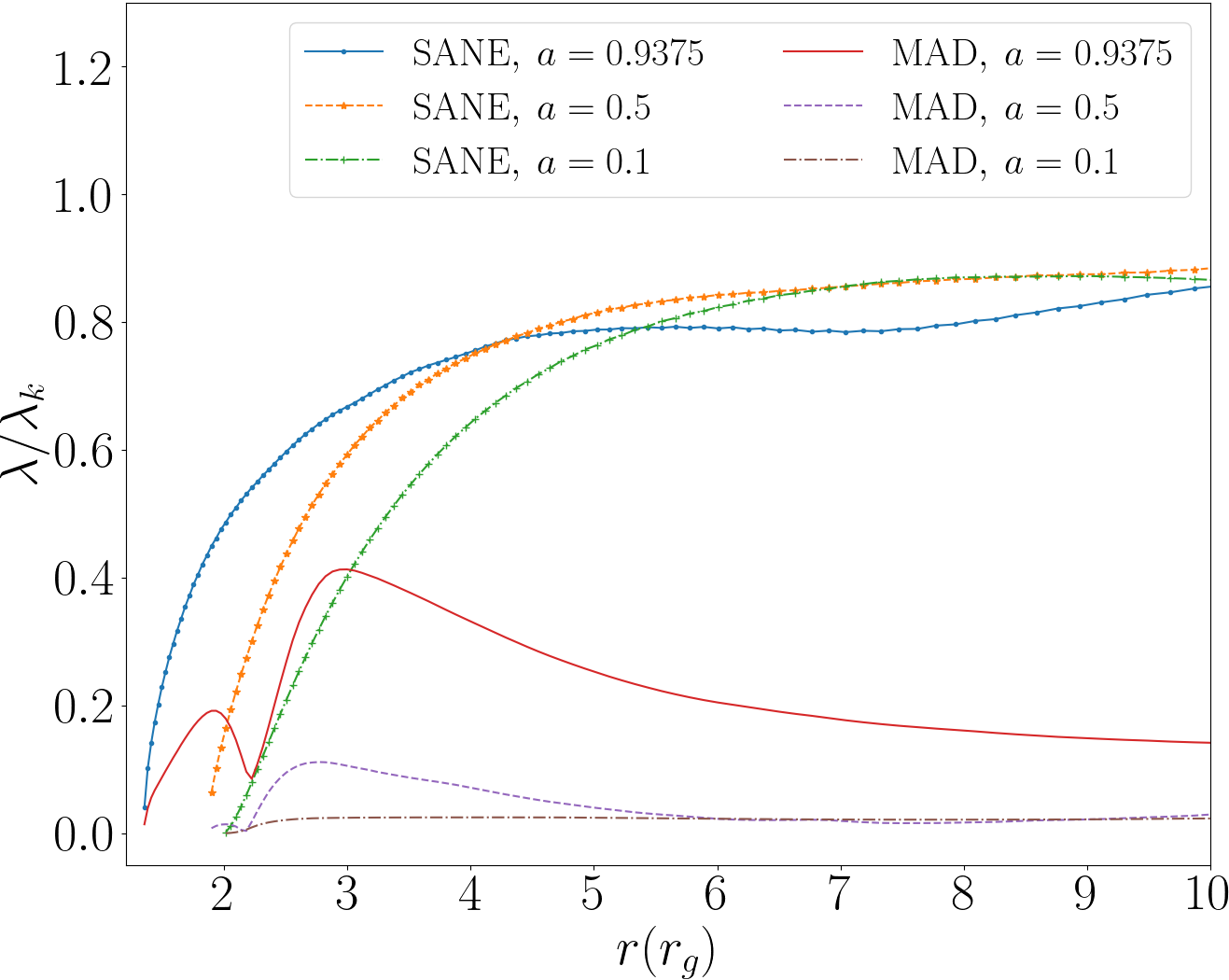}
     \caption{SANE and MAD angular momentum profiles.}
     \label{fsrqllk}
 \end{figure}

The FSRQ spectra also exhibit an optical component. BL Lac spectra, on the other hand, show no such optical spectral features. In Fig. \ref{fsrqllk}, we have shown the ratio of angular momentum of the flow ($\lambda$) to Keplerian angular momentum ($\lambda_k$) \citep{pp02}, 
given by
\begin{equation}
    \lambda_k=\frac{r^2-2a\sqrt{r}+a^2}{\sqrt{r}(r-2)+a},
\end{equation}
where $a$ is the spin of the black hole, $r$ is the distance from the 
black hole, $\lambda=v^\phi r$ where $v^\phi$ is the $\phi$-component of the three-angular velocity in the ZAMO (zero angular momentum observer) frame.

$\lambda/\lambda_k$ for SANE systems is higher than MAD for all black hole spins considered. Moreover, for SANE systems, the $\lambda/\lambda_k$ values approach $1$ away from the black hole. This shows similarities between SANE and the geometrically thin Shakura-Sunyaev disk model \citep{ss73}. As a thin disk is suitable to explain optical emissions from accretion sources, SANE quite corroborates with FSRQs with the assumption of its association of a Keplerian disk away from the black hole. While SANE by itself cannot be a Keplerian disk, its presence seems to be suggesting the existence of an accompanying Keplerian disk. MAD systems, however, have very low $\lambda/\lambda_k$, indicating a radial velocity dominated flow, which is a characteristic feature of the sub-Keplerian advective accretion flow. Several non-thermal spectra have been well explained by the advective disk model.

\subsection{Plasma-$\beta$}

 \begin{figure}
\includegraphics[width=0.48\textwidth]{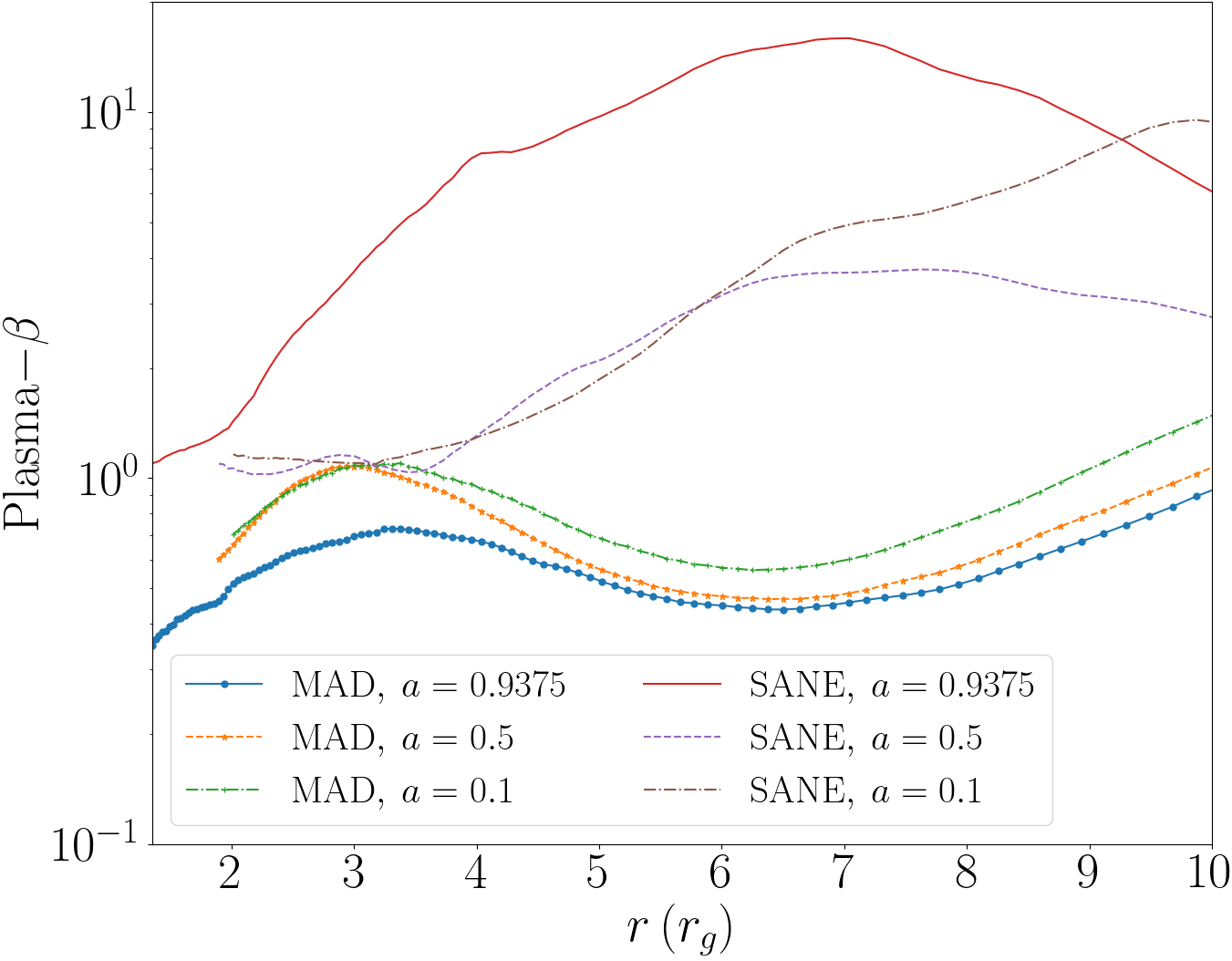}
     \caption{Plasma-$\beta$ profiles for all the simulations considered.}
     \label{pb}
 \end{figure}

MRI is one of the main sources of angular momentum transport in accretion flows. For MRI to work effectively (see, however, \citealt{bgm}, for non-axisymmetric flows), the magnetic fields should not be very high. In other words, the plasma-$\beta$ of the accretion flow should not be very low. It has been argued that a plasma-$\beta\gtrsim 1$ is most effective for MRI to cause effective angular momentum transport. At very high fields (or at plasma-$\beta\lesssim 1$), the fields become too strong to produce other instabilities in the accretion flow.

In Fig. \ref{pb}, we show the disk averaged plasma-$\beta$ profiles of the accretion flow in our simulation domain. As is clearly evident, the plasma-$\beta$ for MAD flows is less than $1$ almost throughout the inflow-outflow equilibrium region. For SANE flows, however, the plasma-$\beta$ is more than $1$. This shows that SANE flows are prone to MRI induced turbulence, leading to angular momentum transport. For MAD flows, on the other hand, the flow is magnetically dominated and the fields are too strong for any MRI to occur. This perfectly corroborates with the result obtained in sec. 4.1, that MAD flows are dominated by large scale magnetic fields, while SANE flows have higher contribution from MRI induced small scale magnetic fields.
 
Thus, the magnetic stresses, $\lambda/\lambda_k$ profiles and plasma-$\beta$ profiles indicate that SANE systems are more disk dominated systems, dominated by MRI induced angular momentum transport, thus making them good candidates for explaining the underlying accretion flow in FSRQs with their spectral features.
MADs exhibit magnetically dominated nature with large scale magnetic field induced angular momentum transport and advective properties. It can thus provide a good explanation for the observational properties of BL Lacs and their corresponding accretion flow characteristics. 

\section{Conclusion}

Our simulation results have shown that hard state ULXs can be interpreted as magnetically arrested sub-Keplerian accretion flows around fast spinning stellar mass black holes. The observational luminosity of hard state ULXs can be understood by considering them as MAD systems. 
Such MAD accretion flows are known to have more than 100\% efficiencies for fast spinning black holes \citep{rn2011}. This is because the flow taps the magnetic energy present in the system due to the magnetically dominated nature of the energy flux. This effect, paired with extracting the rotational energy of the black hole via the Blandford-Znajek mechanism, leads to such high efficiencies in MAD systems. This makes MAD a very promising candidate for explaining the accretion flow properties which lead to the peculiar observed luminosities for hard state ULXs.

We have also shown that the FSRQ/BL Lac dichotomy can be explained by considering them to be accreting systems around supermassive black holes with different magnetic properties. Their debeamed luminosities calculated by BL19 also enable us to interpret their spectral behaviour as a result of the spin of the central rotating black hole. FSRQs can be explained as SANE systems around slow to intermediate rotating black holes while HSP to ISP BL Lacs are MAD accretion systems around slow to fast rotating black holes respectively. Due to the observational similarities between LSP BL Lacs and FSRQs, we interpret LSP BL Lacs as SANE accretion systems around fast rotating black holes.

The nature of the angular momentum transport can then be understood. BL Lacs have pre-dominantly large scale magnetic fields, driving the outflow and angular momentum transport. In the case of FSRQs, small scale magnetic fields and MRI driven turbulence drives the dynamics of the system. The magnetic stress profiles along with angular momentum and plasma-$\beta$ profiles show that FSRQs are more disk-dominated systems which also explains the optical component present in their observed spectra. Similarly, the magnetically dominated nature of BL Lacs explains the dominant non-thermal characteristics in their spectra.

Note, however, for blazars to explain FSRQ and BL Lacs, we have considered a range of black hole spin. Since non-spinning black holes do not seem to be very natural, we have considered the $a=0.1$ case as a proxy for slow spinning black holes. For ULXs, on the other hand, our main focus is on high spinning highly magnetised systems which, as shown by ULX19, are capable of producing high luminosities. We have considered the Schwarzschild case as well for ULX to show theoretical comparison of powers with the change of spin.

Our simulation is axisymmetric, i.e. we have practically carried out 2.5-dimensional GRMHD simulations. Although these simulations have helped us in understanding the dynamics of magnetised accretion systems, they fail to capture the effect of non-axisymmetric phenomena like turbulence, actual flux eruption events etc. In the future, we plan to do full scale three-dimensional simulations of magnetised advective accretion flows to understand the effect of these non-axisymmetric phenomena on different properties of the system like the outflow power, distributions of system variables, like, magnetic fields, accretion rate, density etc. In addition to this, we also plan to study the spectral properties of these accretion flows by including thermal and non-thermal cooling to the system. This will lead to more observationally relevant results and will enable us to compare our synthetic spectra with observationally obtained spectra of ULXs and blazars. This will lead to a better understanding of the outflow production mechanism in these systems. ULXs are known to have a missing iron line in their spectra, while iron lines are a very prominent feature in X-ray binary spectra. Analysing the simulated spectra for ULXs and then comparing it with their observational spectra can shed light on the absence of the iron line. A more rigorous modeling of blazars with BLR (broad line region) clouds and an ambient medium can lead to a better understanding of the EC phenomena observed in the spectra of FSRQs. A comprehensive analysis of radiative flux received along different lines of sight combined with spectral analysis can lead to a better interpretation of the underlying accretion flow in these systems.
\section*{Acknowledgment} 
MP acknowledges the Prime Minister’s Research Fellows (PMRF) scheme for providing fellowship. BM would like to acknowledge the project funded by SERB, India, with Ref. No. CRG/2022/003460, for partial support to this research.
We thank Arif Babul (UVic), Koushik Chatterjee (UMD), Christian Fendt (MPIA, Heidelberg), Ramesh Narayan (Harvard) and Bhargav Vaidya (IIT Indore) for helpful discussions. We also thank Hector Olivares (Aveiro University) for helping with certain technical problems related to BHAC. Finally, thanks are due to the referee for a thorough review with useful comments.

\bibliography{Reference.bib}
\bibliographystyle{aasjournal}

\end{document}